\documentclass[preprint]{aastex631}

\shorttitle{Radio Emission from UV Cet}
\shortauthors{Bastian, Cotton, \& Hallinan}

\begin{document}

\title{Radio Emission from UV Cet: Auroral Emission from a Stellar Magnetosphere}

\correspondingauthor{T. S. Bastian}
\email{tbastian@nrao.edu}

\author[0000-0002-0713-0604]{T. S. Bastian}
\affiliation{National Radio Astronomy Observatory, 520 Edgemont Road, 
Charlottesville, VA 22903 USA}

\author[0000-0002-0713-0604]{W. D. Cotton}
\affiliation{National Radio Astronomy Observatory, 520 Edgemont Road, 
Charlottesville, VA 22903 USA}

\author[0000-0002-0713-0604]{G. Hallinan}
\affiliation{Department of Astronomy, California Institute of Technology, Pasadena, CA 91125, USA}

\begin{abstract}
\noindent The archetypical flare star UV Cet was observed by MeerKAT on 5-6 October 2021. A large radio outburst with a duration of $\sim\!2$ hr was observed between 886-1682 MHz with a time resolution of 8s and a frequency resolution of 0.84 MHz, enabling sensitive dynamic spectra to be formed. The emission is characterized by three peaks containing a multitude of broadband arcs or partial arcs in the time-frequency domain. In general, the arcs are highly right-hand circularly polarized. During end of the third peak, brief bursts occur that are significantly elliptically polarized. We present a simple model that appears to be broadly consistent with the characteristics of the radio emission from UV~Cet. Briefly, the stellar magnetic field is modeled as a dipole aligned with the rotational axis of the star. The radio emission mechanism is assumed to be due to the cyclotron maser instability where x-mode radiation near the electron gyrofrequency is amplified.  While the elliptically polarized bursts may be intrinsic to the source, rather stringent limits are imposed on the plasma density in the source and along the propagation path. We suggest that the elliptically polarized radiation may instead be the result of reflection on an over-dense plasma structure at some distance from the source. Radio emission from UV~Cet shares both stellar and planetary attributes. 
\end{abstract}


\section{Introduction}

The dwarf star UV~Cet is the archetype of flare stars, producing frequent outbursts in the optical (O), ultraviolet (UV), soft X-ray (SXR), and radio wavelength bands. It is the secondary in UV~Ceti (Gleise~65 = Luyten~726-8), a wide binary system ($P=26.28$ yrs); the primary is BL~Cet, also a flare star. At a distance of just 2.68~pc UV~Ceti has been a subject for studies of the magnetic field and magnetic activity on both stars. Despite the fact that UV~Cet and BL~Cet are similar in many respects \citep{Kervella2016} -- in spectral type (M6V and M5.5V, respectively), rotation ($v\sin i\sim 30$ km-s$^{-1}$), mass ($\approx 0.12$ M$_\odot$), and radius ($\approx 0.16$ R$_\odot$) -- they appear to have quite different magnetic fields \citep{Kochukhov2017}. Both stars have strong average surface magnetic fields $\langle B\rangle > 4$ kG but the magnetic field of BL~Cet is complex (multipolar). In contrast, the magnetic field of UV~Cet is dominated by a strong axisymmetric dipolar field. The dramatic difference between the magnetic fields of two otherwise similar stars challenges our understanding of dynamos in fully convective objects \citep{Kochukhov2017a, Shulyak2019, Kochukhov2021}. Magnetic activity on the two stars has likewise attracted renewed interest. With the discovery of multitudes of extrasolar planets, e.g. \citet{Winn2015}, with many orbiting late-type dwarf stars, the question of whether stellar activity and stellar space weather are significant factors in determining the formation, evolution, and habitability of exoplanets \citep{Osten2018} has come to the fore. 

At radio wavelengths, UV~Cet poses both puzzles and opportunities. With the discovery of symmetrical, large-scale radio-emitting structures using VLBI techniques \citep{Benz1998} it has long been suspected that it may have an extended magnetosphere, consistent with work cited above. Recent radio observations have shown that radio outbursts near a frequency of 1~GHz in fact recur on UV Cet \citep{Zic2019} with the star's rotation period of 5.45~hrs, again consistent with the presence of a large-scale organized magnetic field. However, the details of the extended stellar field remain largely unknown but as we show here, radio observations offer the possibility of constraining several aspects of the large scale magnetic field and the plasma environment. 

Radio emission also offers potential opportunities for the detection and exploitation of analogs of solar flares and space weather tracers such as radio bursts of type II (shocks due to fast coronal mass ejections), type III (produced by nonthermal electron beams), and type IV (produced by trapped populations of electrons). While flares are commonly detected on late-type dwarf stars, several searches for space weather tracers at radio wavelengths on late-type dwarf stars \citep{Osten2006,Osten2008,Crosley2016,Crosley2018,Villadsen2019} yielded no compelling detections of type II/III analogs.  \citet{Zic2020} have claimed the detection of a possible analog of type IV emission from Proxima Centauri, however. A barrier to more fully understanding the nature of radio outbursts from flare stars in general and UV~Cet in particular has been uncertainty about the basic emission mechanism(s). Radio bursts have generally been attributed to either plasma radiation or to the electron cyclotron maser instability \citep{Bastian1990a}. The former mechanism is responsible for most solar radio bursts below 1-2~GHz \citep{Bastian1998} while the latter is highly relevant to planetary auroral emissions \citep{Zarka1998, Badman2014}. In recent years, evidence has accumulated that favors coherent auroral radio emission from UV~Cet \citep{Benz1998, Schrijver2009, Lynch2017, Villadsen2019, Zic2019}, at least for radio emission at decimeter and meter wavelengths. The observations presented here strongly favor this interpretation. 

In this paper we present sensitive, wideband, high-resolution dynamic spectra of a powerful radio outburst from UV~Cet. The observations and data reduction are described in \S2. In \S3 we present the observations in detail. In \S4 we develop a simple model of a rigidly rotating magnetosphere in which the cyclotron maser instability drives the observed emission, and show that it can qualitatively account for many features of the observed spectrum. We then consider the aspects of the observed polarization properties of the emission, showing that they may be the result of propagation effects rather than intrinsic to the source. We conclude in \S5. 


\section{Observations and Data Reduction}

MeerKAT is a Fourier synthesis radio telescope in the Northern Cape Province of South Africa \citep{Jonas2016}. It comprises 64 offset Gregorian antennas, each with an effective diameter of 13.5 m, that sample antenna baselines ranging from 29 m to 7.7 km. MeerKAT currently supports observations in the L band (856-1712 MHz) and the UHF band (544-1088 MHz). The observations of UV Ceti reported here were performed by MeerKAT from approximately 20:20 UT on 2021 October 5 to 01:50 UT on 2021 October 6 using the L band receiving system and 58 of the 64 antennas. They were part of proposal code SCI-20210212-TB-02. The observations employed the “wideband coarse” observing mode where 4096 channels were observed across the band, each channel being approximately 209 kHz in width.  The native polarization of the MeerKAT antenna feeds is linear and all correlation products – XX, YY, XY, and YX – were recorded with a time resolution of 8 s.  The calibrators J1939-6342 and J0408-6545 were observed for 10 min. at the beginning and end of the observation to establish flux and bandpass calibration; J0240-2309 was used as the gain calibrator and was observed for 3~min every $\sim\!30$  min. 

Data processing was performed using the Obit package \citep{Cotton2008}. The parallel hand calibration and editing largely followed the
development in \citet{Mauch2020} except that the data were not time averaged beyond the 8 second samples of the observations. The calibration procedure was modified to allow polarization calibration. The data did not contain a polarized calibrator and so polarization calibration followed \citet{Plavin2020}. This calibration uses the ``noise diode'' calibration at the beginning of each observing session to determine the bulk of the X-Y phase function together with calibration tables derived from other, better calibrated datasets. The X-Y gain ratios were set using the bandpass calibration on the very weakly polarized source J1939-6342 and subsequent gain calibration solved for Stokes I. After correction for on--axis instrumental polarization, the data were transformed to a circular basis (i.e. RR, LL, RL, LR) and averaged in frequency to a resolution of 0.84 MHz (see below). The flux density scale is based on the spectrum of J1939-6342 \citep{Reynolds1994} given by $\log(S) = -30.7667 + 26.4908 \log\bigl(\nu\bigr) - 7.0977 \log\bigl(\nu\bigr)^2 + 0.605334 \log\bigl(\nu\bigr)^3$, where $S$ is the flux density (Jy) and $\nu$ is the frequency (MHz).

The Stokes I data were then imaged using the Obit wideband imager MFImage including self-calibration. MFImage uses faceting to correct for the noncoplanarity of the sky and uses multiple constant fractional bandwidth subbands (here 5\%) which are imaged independently and deconvolved jointly. Imaging fully covered to a radius of 1.2$^\circ$ with outliers within 1.5$^\circ$ estimated to be above 1 mJy from the NRAO VLA Sky Survey \citep{Condon1998}, or the Sydney University Molonglo Sky Survey \citep{Mauch2003} catalogs.  CLEANing proceeded to a residual level of 100 $\mu$Jy beam$^{-1}$  and CLEANed 1.77 Jy of emission. Fig. 1 shows the inner 1$^\circ\times1^\circ$ of the map in which dozens of background sources are present.   The presence of UV~Ceti near the center of the image is obvious. Its time variability resulted in unCLEANable sidelobes that appear as artifacts in the map. 

The components of the Stokes I CLEAN model further than 20'' from the position of UV Ceti were subtracted from the visibillity data so that
only the response of UV~Ceti remained.  Further data processing was performed in the Astronomical Image Processing System (AIPS) and Interactive Data Language (IDL). AIPS does not form Stokes parameters from correlation products using a linear polarization basis -- hence, the conversion to a circular basis. Moreover, the IDL utilities used were originally written for use with the AIPS visibility data format. The spectrally-averaged and visibility-subtracted data were phase-shifted to place UV Ceti precisely at the phase center and further averaged to 240 frequency channels with a resolution of 3.34 MHz per channel.  Maps of the model-subtracted data were formed in Stokes I, Q, U, and V for the inner square degree of field of view to check for artifacts resulting from inadequate background source subtraction, polarization leakage, or beam squint. Stokes I represents the total intensity, Stokes V represents the circularly polarized emission, Stokes Q and U represent linearly polarized emission. Although linearly polarized background sources were present, the strongest produced only 1.4~mJy of polarized flux. It was at an angular distance of 21' from the target source and contributed negligible sidelobes. Non-stellar background sources typically show negligible circularly polarized radiation and squint-induced circularly polarized artifacts were well suppressed. We conclude that our calibration and model subtraction procedures produced a visibility data base that was largely free of polarization leakage or other artifacts. 

As an unresolved point source, dynamic spectra of UV~Ceti could be easily formed by computing, for each integration time, frequency channel, and correlation product, the mean of the real part or imaginary part of the complex visibility on all antenna baselines. This was done to produce spectra of the RR, LL, RL, and LR correlation products, as well as iRL and iLR.  From these, dynamic spectra of the Stokes polarization parameters were formed as $\rm I = (RR+LL)/2$, $\rm V = (RR-LL)/2$, $\rm Q = (RL+LR)/2$, and $\rm U =-i(RL-LR)/2$.  The total linearly polarized emission is $\sqrt{Q^2+U^2}$ and the electric vector position angle (EVPA) of the linearly polarized component is $\phi=\tan^{-1}(Q/U)/2$. We now describe key results from these spectra. 


\begin{figure}
\begin{center}
\includegraphics[width=5in]{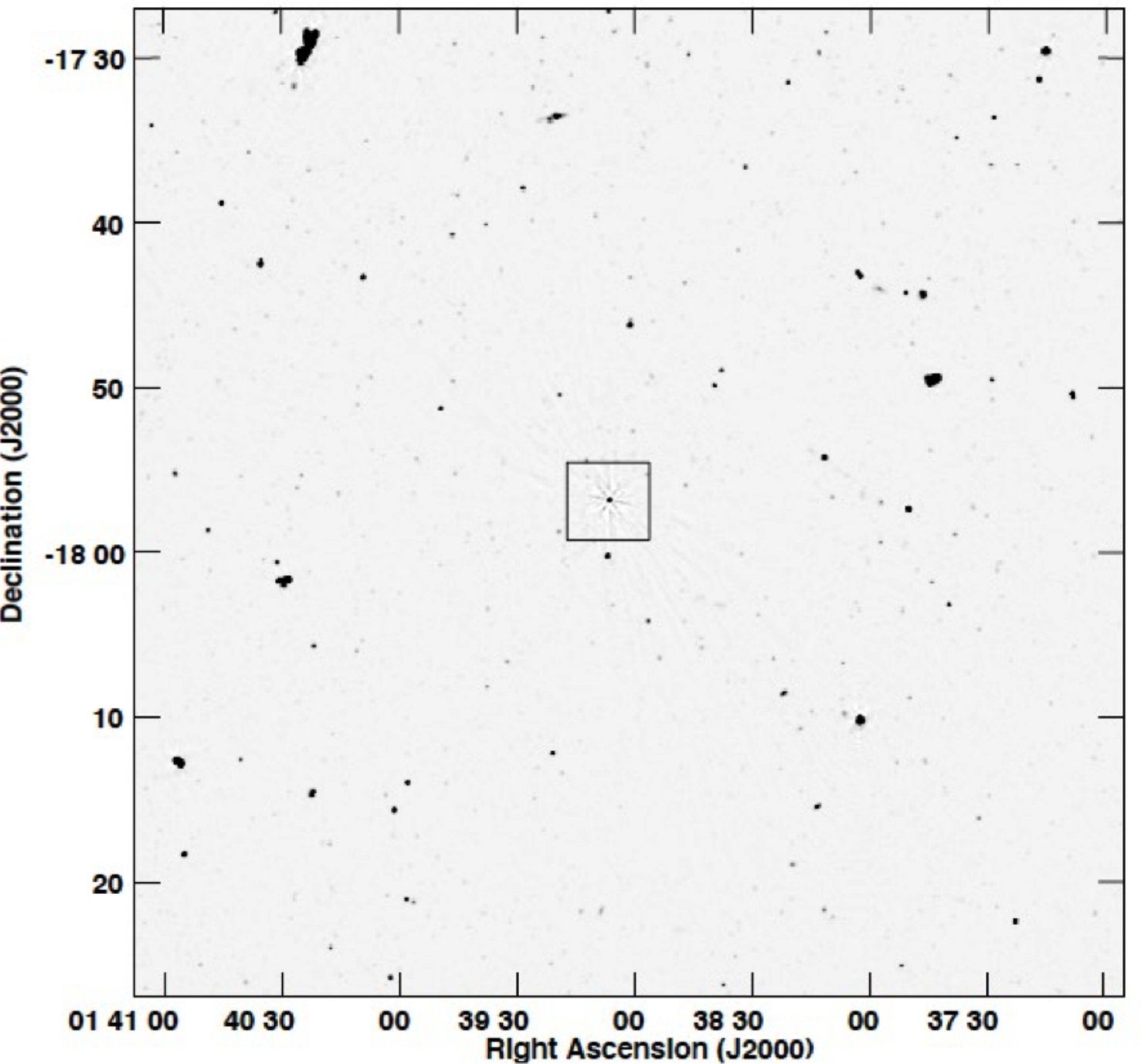}
\caption{MeerKAT map of the field containing UV~Ceti (box). The field of view has been restricted to the central $1^\circ\times 1^\circ$. It has been clipped below -0.2~mJy/bm and above 2~mJy/bm to better reveal the multitudes of background sources. The sidelobe structure around UV~Ceti is due to its time variability during the 5.5~hr observation. }
\end{center}
\end{figure}

\begin{figure}
\begin{center}
\includegraphics[width=6.5in]{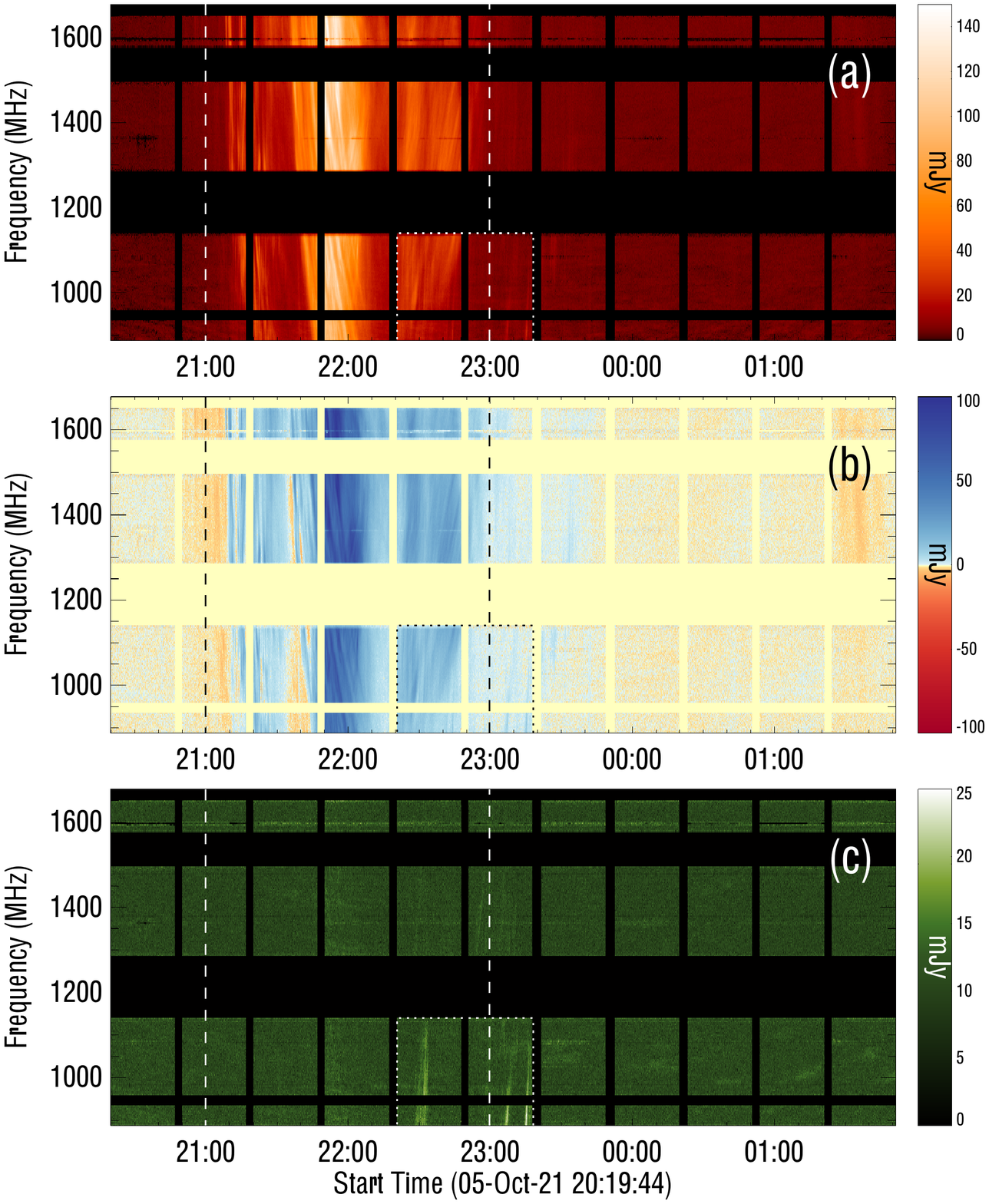}
\caption{Overview of the radio emission from UV~Ceti during the course of just over one rotational period. a) The dynamic spectrum of Stokes I, showing the total density  as a function of time and frequency. The vertical gaps are times that the gain calibrator was observed. The horizontal gaps represent frequencies corrupted by RFI that were flagged out of the dataset. b) The dynamic spectrum of Stokes V. Blue indicates RCP emission and orange to red indicates LCP emission. c) The dynamic spectrum of the linearly polarized flux density $\sqrt{Q^2+U^2}$.  The vertical dashed and dotted boxes indicate time and frequency ranges that are considered in greater detail in the next section. }
\end{center}
\end{figure}



\section{Results}

An overview of the radio emission from UV~Cet is shown in Fig.~2 for the 5.5~hr duration of the observation, just over one full rotation of the star ($P_{rot} = 5.45$~hrs). The dynamic spectrum of Stokes I is shown in panel (a), that of Stokes V (circularly polarized radiation) is shown in panel (b); and the linearly polarized radiation, corrected for noise bias \citep{Mueller2017} is shown in panel (c). The spectra are scaled by the cube root of the flux density in each case to better emphasize faint emission. The bandpass-averaged light curve of Stokes I is shown in the top panel of Fig.~3. A radio outburst of $\sim\!2$ hrs duration dominates the emission, reaching a peak average flux density 105~mJy before returning to a quiescent level of a few mJy. We note that the maximum in the Stokes I dynamic spectrum is 147~mJy at a frequency of 1392 MHz at 21:52 UT. The standard deviation of the Stokes I and V spectra is $\approx 0.7$ mJy per channel per integration time.


\begin{figure}
\begin{center}
\includegraphics[width=6.5in]{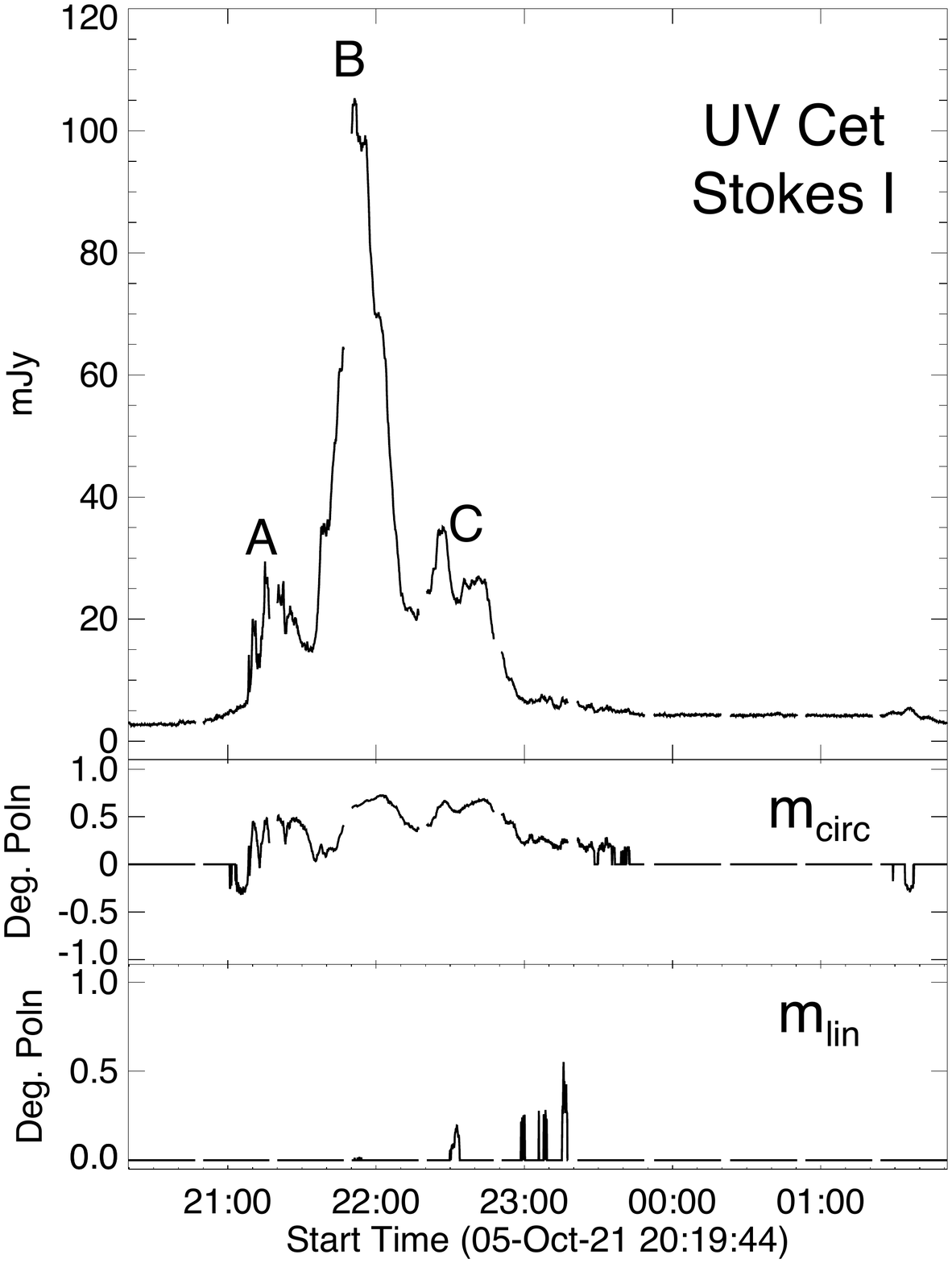}
\caption{Top: The light curve of the radio emission from UV~Cet integrated over the frequency bandpass. Gaps are times when the gain calibrator was observed. Middle: the fractional circular polarization. Bottom: The fractional linear polarization. In the case of the Stokes I and Stokes V the data are averaged over the entire bandpass. Since the linearly polarized emission is confined to lower frequencies the average is performed over a sub-band from 960-1127 MHz. }
\end{center}
\end{figure}

The event is characterized by three broad peaks that we label A, B, and C in Fig.~3.  The emission spans the entire bandwidth in Stokes I and V although drifting substructures with narrow bandwidths are clearly visible within each peak.  The degree of circular polarization $m_{circ}=V/I$ is shown in the middle panel of Fig.~3 for those points for which the Stokes I flux density is $>5$ mJy. The outburst is strongly right-hand circularly polarized (RCP) throughout the outburst, with the average peak degree of polarization ranging from of 0.5 during peak A to more than 0.7 (peaks B and C). As best seen in Fig.~2, however, the emission is weakly to moderately left-circularly polarized (LCP) prior to several sub-peaks during peak A and a weak LCP burst also occurs near the end of the observations at 01:40 UT. The degree of linear polarization is $m_{lin}=(Q^2+U^2)^{1/2}/I$ is shown in the bottom panel of Fig.~3.  The outburst shows no significant linearly polarized emission until the declining phase of peak C, where linearly polarized features below $\approx\!1120$ MHz are seen as well as faint stria near 1400~MHz. 
We now discuss the dynamic spectrum and the polarization properties of the outburst in greater detail. 
\pagebreak

\subsection{The Dynamic Spectrum}

Fig.~4 shows a detail of the Stokes I dynamic spectrum for the 2~hr period of outburst emission in the top panel. The times corresponding to Peaks A, B, and C are again indicated. The Stokes I and V spectra reveal complex structure in the time-frequency domain. Particularly striking are arcs of emission in peak B. To better show the substructure in the Stokes I dynamic spectrum we show two representations of the spectrum in Fig.~5. In the top panel we have applied a modified unsharp mask in the time dimension to emphasize low-contrast features. In the lower panel the dynamic spectrum has been passed through a Sobel edge-ehancement filter. 
In peak A, the discrete substructures show little change in frequency with time although the feature to the left of label ``a'' drifts from higher to lower frequencies with time and the overall envelope of peak A likewise first appears at high frequencies and then drifts down to lower frequencies.  Peak B shows more coherent substructures that clearly trace arcs, partial arcs, or bundles of arcs that drift from high to low frequencies from approximately 21:40-22:00 UT. Features to the right of labels $b_1$, $b_2$, and $b_3$ show that the arcs in peak B indeed display curvature in the sense of increasing drift rates with decreasing frequency. These particular features drift at approximately -7~MHz-s$^{-1}$, -2~MHz-s$^{-1}$, and -1~MHz-s$^{-1}$, respectively. At later but overlapping times, from 21:50-22:10 UT, substructures in peak B show the opposite sense of drift, from low to high frequencies. Particularly notable are features near $b_4$ between roughly 1100-1400~MHz for which apparent frequency drift rates are $\approx\!0.5$ MHz-s$^{-1}$. Finally, in peak C, distinct substructures are sparser in number but, like the trailing part of peak B, the substructures drift from low to high frequencies (e.g., $c_2$) with time between roughly 950-1500 MHz, again at rates $\approx 0.5$ MHz-s$^{-1}$ although there are also nearly vertical structures ($c_1$) present that show rapid change in frequency with time. 

\begin{figure}
\begin{center}
\includegraphics[width=6.5in]{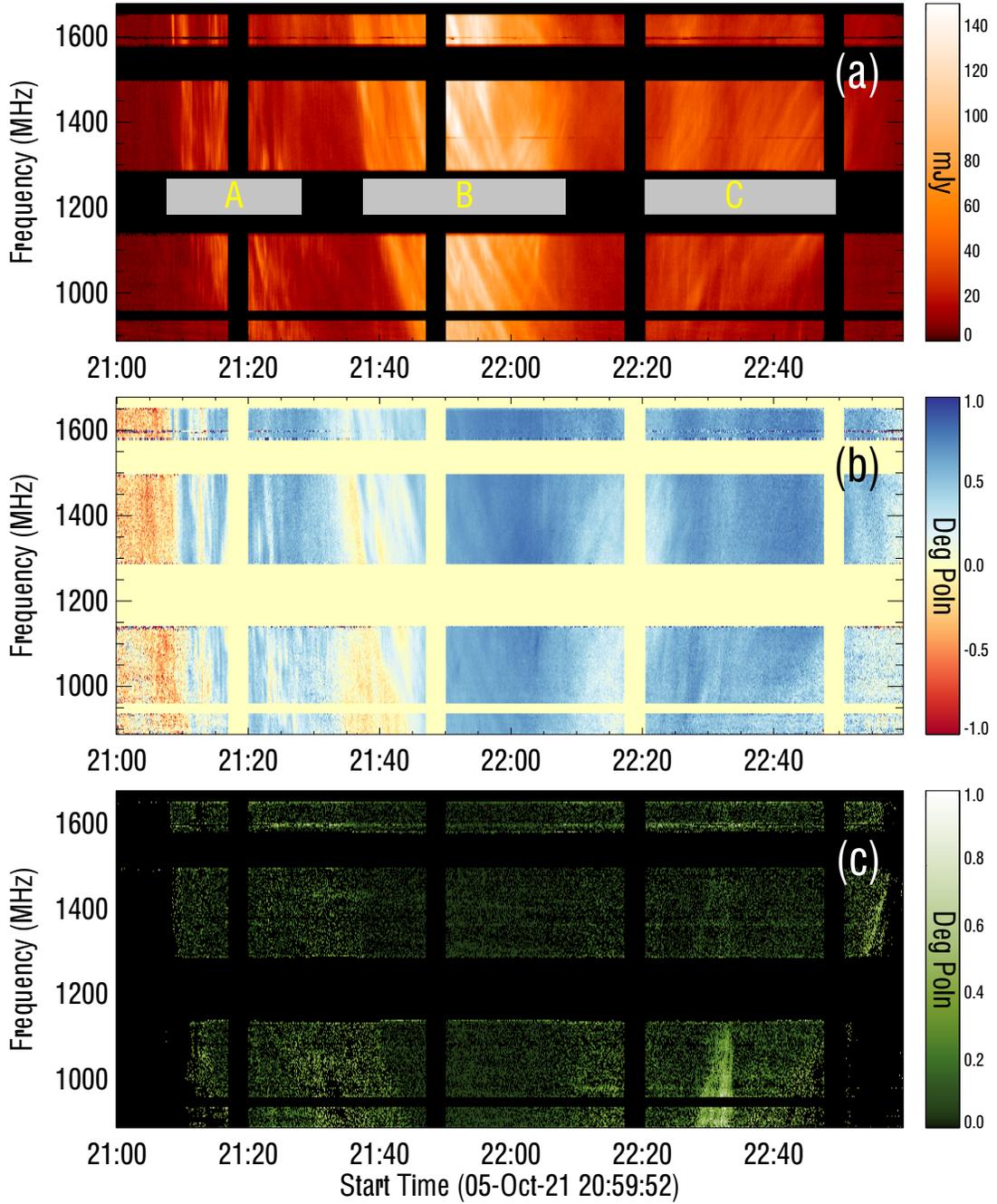}
\caption{Detail of the dynamic spectra between 21:00-23:00 UT. Top: the dynamic spectrum of Stokes I; Middle: dynamic spectrum of the degree of circular polarization $m_{circ}$; Bottom: dynamic spectrum the degree of linearly polarized emission $m_{lin}$. Note that $m_{lin}$ is quite noisy where the total intensity emission is low. The color table has been scaled to the cube root of the brightness in all panels. }
\end{center}
\end{figure}

It is difficult to characterize the properties of individual arcs or partial arcs because in many cases they overlap or appear to comprise multiple structures. Moreover, given the integration time of 8~s, it is not clear that these structures are fully resolved in time. However, they may be adequately resolved in frequency; at least, at this time resolution. Spectra examined with a frequency resolution of 0.84~MHz did not show narrower-bandwidth structures than those seen in the data averaged to channel widths of 3.34~MHz. Bandwidths of discrete structures range from $\sim\!10$ MHz to perhaps as much as $\sim\!100$ MHz. Discrete features have durations that appear to be close to the integration time of 8~s to no more than 1~min at a given frequency. A detailed characterization of the frequency bandwidths and durations of discrete emission features is beyond the scope of this paper, however. 

To summarize, the radio outburst observed on UV~Cet comprises multitudes of narrowband substructures that drift in frequency with time, from high to low frequencies during the first half of the outburst and then from low to high frequencies during the second half of the burst.  Fast-drift structures are seen in peak A although the envelop of emission drifts from high to low frequencies. Peak B displays coherent arcs of emission that first drift from high to low frequencies and then drift from low to high frequencies. Peak C shows both fast-drift and slow-drift components that drift from low to high frequencies. 

\begin{figure}
\begin{center}
\includegraphics[width=6.5in]{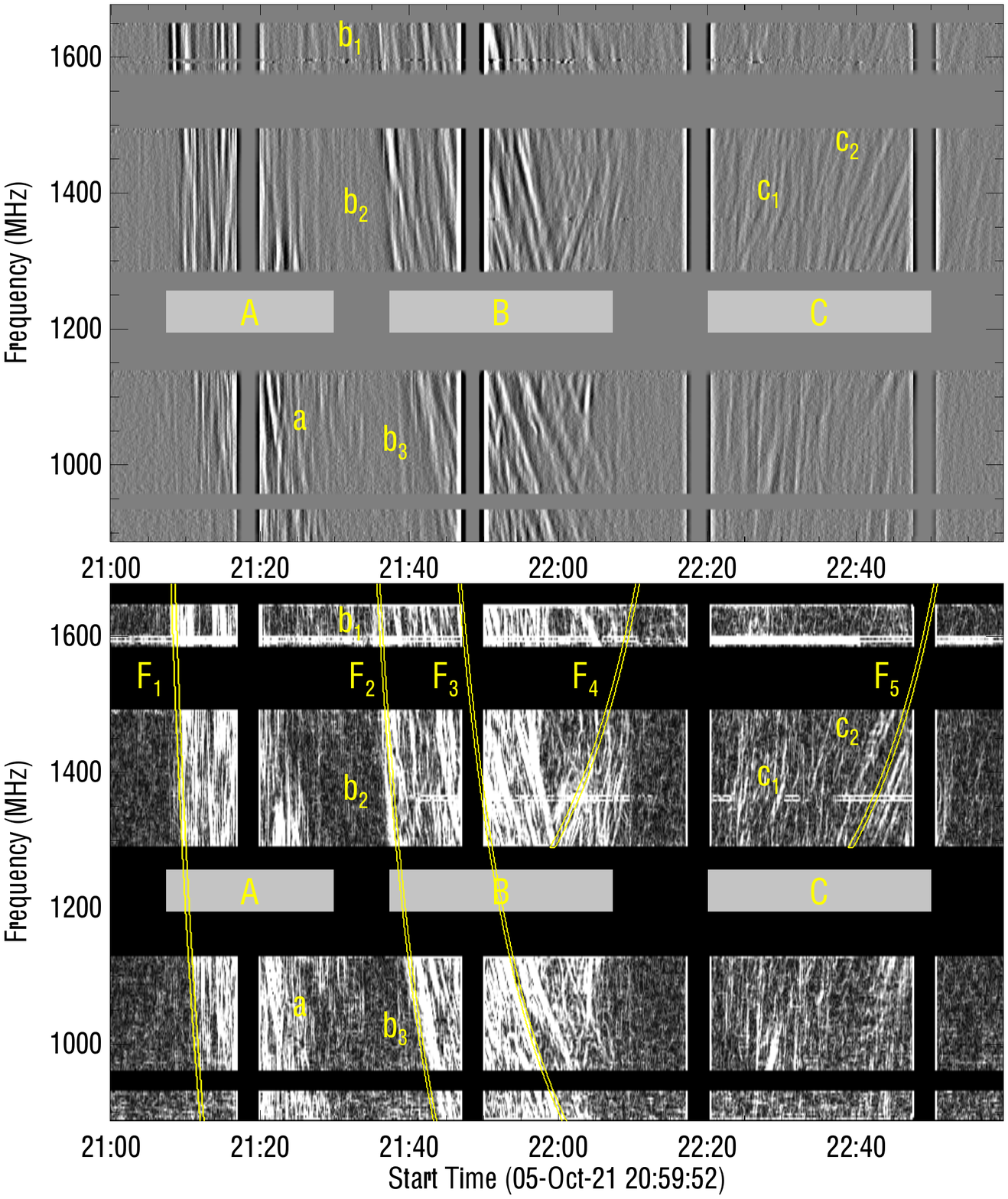}
\caption{Top: Detail of the Stokes dynamic spectrum of the radio outburst from UV~Cet that has been unsharp masked in the time domain. Bottom: the same, but edge-enhanced by a Sobel filter. The bottom panel also shows notional fits of active field lines to specific features. Details are given in \S4.4. }
\end{center}
\end{figure}

We briefly compare our observations with those reported by \citet{Zic2019} who observed UV~Ceti with the Australian Square Kilometer Array Pathfinder (ASKAP; \citet{Hotan2021}). They observed UV~Ceti for $\approx\!10$ hrs on two different days, separated by about 6~months, with a similar integration time (10~s) and spectral resolution (4~MHz) to the observations reported here. However, they observed a smaller bandwidth (744-1032 MHz) with lower sensitivity (7.4 and 9.7 mJy on two different observing epochs) than the MeerKAT observations.  A key finding by \citet{Zic2019} was that radio outbursts on UV~Cet recur with the rotational period of the star $P_{rot}=5.45$ hr. Moreover, even substructures and their drift rates recur from one rotation to the next and, while substructures differ, drift rates remain comparable even between the two observing epochs. Zic et al. find drift rates for the substructures reported in the 744-1032~MHz band of  about 1.4 MHz s$^{-1}$, which we do not view as being inconsistent with those we find in the lower frequency portion of the MeerKAT spectrum. 


\subsection{Polarization}

Figs.~2, 3, and 5 show that the radio emission from UV~Cet is predominantly RCP throughout the event. For peak A, the fractional polarization is broadly distributed between 0.1-0.6 RCP; that of peak B shows broad distribution from 0.1-0.5 but peaks strongly at 0.62 with values extending above 0.75 RCP; and the fractional polarization in peak C is strongly peaked at 0.57 with values also extending above 0.75 RCP.
However, weak LCP features do occur prior to peak A and in the interstices of RCP-polarized substructures in peaks A and B. The Stokes I flux density is low at these times ($\sim\!5-10$ mJy), and while the degree of polarization is $\sim\!0.2$ LCP the uncertainty is comparable in magnitude. By way of comparison, for the quiescent period from 00:00-01:00 UT, the average Stokes I flux density is 4.2 mJy and the average Stokes V flux density is $50\pm 7$ $\mu$Jy, yielding a formal degree of circular polarization of $0.01$. We do not have confidence that our polarization calibration is accurate to this level and simply conclude that the degree of circular polarization during quiescence is ``very low". 

Fig.~6 shows a detail of Fig.~2 during the trailing part of peak C. The panels show dynamic spectra for roughly 57~min starting at 22:40~UT for Stokes I, Stokes V, the bias-corrected linearly polarized emission, and the EVPA.  Particularly interesting is the presence of linearly polarized components, most clearly seen at frequencies $<1120$~MHz or so. The first (labeled $l_1$ in Fig. 6), at $\approx$22:30~UT is of 7 min duration at 900 MHz whereas the other two components ($l_2$ and $l_3$, at approximately 23:06 and 23:14~UT, have durations of $\approx\!2$ min. There is also a faint ($<5$ mJy) linearly polarized feature $\sim\!10$ min before $l_2$ as well as faint striations between 1300-1500~MHz (not shown) at these times. We do not discuss these quantitatively, however.  The mean fractional circular polarization of peaks $l_1$, $l_2$, and $l_3$ is $m_{circ}\approx 0.53$, 0.43, and 0.43, respectively, while the mean fractional degree of linear polarization for each peak is $m_{lin}\approx 0.1, 0.56$, and 0.59, respectively. These result in a mean fractional polarization $m_{tot}=(Q^2+U^2+V^2)^{1/2}/I$ of 0.54, 0.7, and 0.73 for the three peaks; i.e., the peaks are elliptically polarized -- two of them are highly elliptically polarized.  Inspection of Fig.~6c shows that the polarized intensity of $l_2$ and $l_3$ declines with increasing frequency. It is less clear whether this is the case for $l_1$. Interestingly, the degree of linear polarization is relatively constant with frequency. Regarding the EVPA we find no evidence that it changes significantly as a function of frequency within a given peak but the signal-to-noise ratio is not high. It does, however, change by $22^\circ\pm 10^\circ$ between peak $l_1$ and peaks $l_2$, $l_3$.  It is also interesting to note that the linearly polarized peak $l_1$ corresponds to a diminution in the flux density of both Stokes I and V at the same time and frequency range (Fig.~6) .


\begin{figure}
\begin{center}
\includegraphics[width=6.5in]{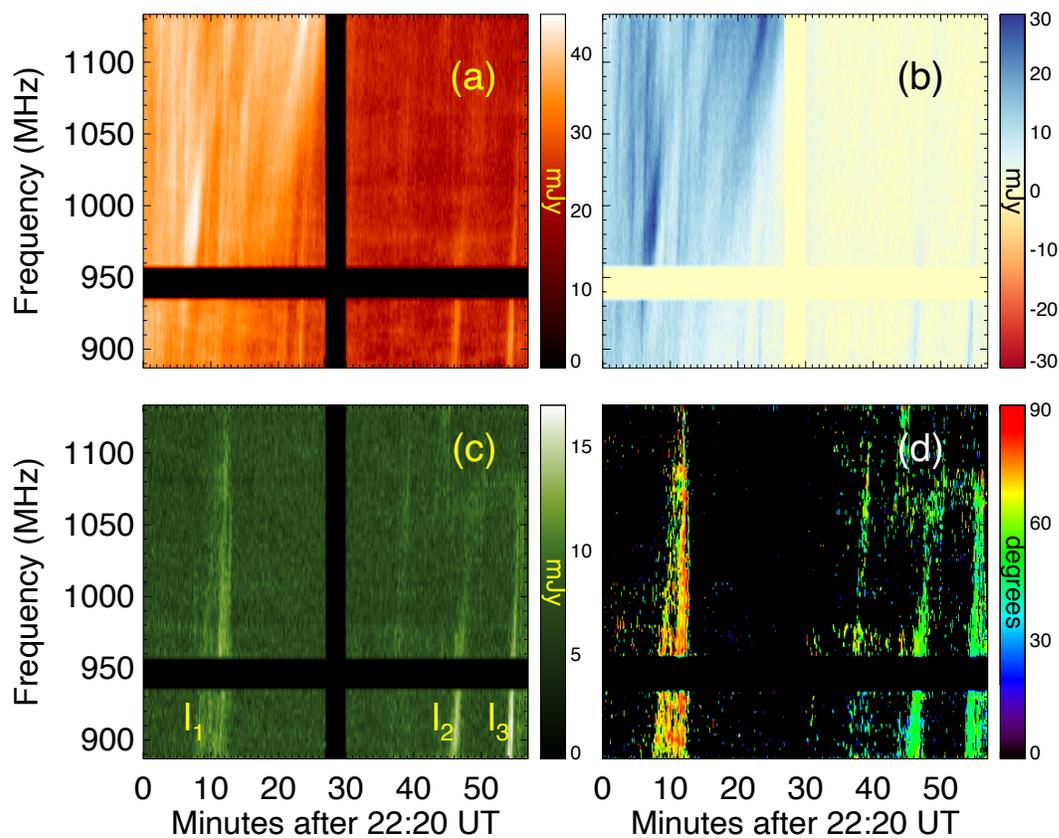}
\caption{Detail of the polarized emission between 21:00-23:00 UT. (a) the Stokes I spectrum; (b) the Stokes V spectrum; Bottom: the linear polarization spectrum; (d) the EVPA of the linearly polarized emission. }
\end{center}
\end{figure}

The MeerKAT observations of polarized emission are consistent with those reported by \citet{Zic2019}. They, too, find that radio outburst emission is strongly RCP, with mean fractional circular polarization of 0.7. No LCP components are reported. They also report the first detections of a linearly polarized component from UV~Cet in this frequency band on both epochs it was observed. However, the fractional degree of linearly polarization is similar to burst $l_1$ (Fig.~6), peaking at 0.15. No highly elliptically polarized components are reported that are comparable to $l_2$ and $l_3$.

\section{Discussion} 

In this section we address 1) the nature of the radio outburst and the drifting substructures observed in the dynamic spectra; and 2) the polarization properties of the outburst. We do so within the context of a simple model: a stellar magnetosphere producing auroral emissions due to the cyclotron maser instability.  Several lines of evidence lead us to this model: first, the large scale magnetic field of UV~Cet is dominated by a strong dipole that is nearly aligned with the rotation axis \citep{Kochukhov2017a}. Second, the intense, variable, narrowband, and highly circularly polarized emission is clearly coherent (see arguments presented by \citet{Zic2019} which we do not reiterate here). Finally, the fact that the radio outburst is recurrent on the stellar rotation period suggests it is the result of sustained, but slowly evolving, magnetospheric processes.  In the following subsections we briefly review the cyclotron maser mechanism, outline our simple model of a stellar magnetosphere, and then address the MeerKAT observations. 

\subsection{The Cyclotron Maser Instability}

The cyclotron maser instability (CMI) is recognized as the mechanism responsible for coherent emissions from planetary magnetospheres, including Earth's auroral kilometric radiation and Jupiter's auroral emissions at decameter to hectometer wavelengths \citep{Zarka1998, Badman2014}. It is also relevant to coherent emissions from certain magnetic chemically peculiar stars such as  $\sigma$~Ori~E \citep{Trigilio2004} and, more recently, has been recognized as operative on brown dwarfs \citep{Hallinan2015, Pineda2017}. CMI emission has long been suspected of being a possible mechanism for coherent radio outbursts on flare stars \citep{Melrose1982, Bastian1987, Bastian1990}. That suspicion further narrowed to suggestions that, at least for late-type dwarfs, auroral CMI emission is likely relevant \citep{Hallinan2008, Schrijver2009}, including UV~Cet \citep{Bingham2001, Lynch2017, Zic2019}.  With the MeerKAT observations of UV~Cet presented here, and those from ASKAP presented by \citet{Zic2019}, we believe the data strongly favor CMI emission from a large scale magnetospheric configuration, analogous in some respects to Jupiter's auroral emissions, albeit at frequencies that are nearly two orders of magnitude higher. 

The CMI mechanism is a maser in the sense that resonant electrons directly amplify electromagnetic waves via negative absorption. In a classical astrophysical maser, a population inversion in the energy levels of a particular molecule (OH, H$_2$O, SiO) can occur via a pumping mechanism, providing the source of free energy.  Stimulation of a transition to a lower energy state determines the frequency of emission. For the CMI, the analog to a transition frequency is the Doppler-shifted electron gyrofrequency, and the analog to the population inversion as the source of free energy to drive the instability is an anisotropy in the electron momentum distribution function. Consider electromagnetic waves with an angular frequency $\omega$ and a wave vector ${\bf k}$ propagating through a population of fast electrons characterized by a momentum distribution function $f({\bf p)}$ where ${\bf p}=\gamma m{\bf v}$, $\gamma$ is the Lorentz factor, and ${\bf v} $ is the velocity vector. The velocity vector can be expressed in terms of components parallel and perpendicular to the magnetic field, $v_{||}$ and $v_\perp$.  The condition for resonance between electromagnetic waves at harmonic $s$ of the electron gyrofrequency $\Omega_{Be}=eB/m_ec$ is given by $\omega=s\Omega_e/\gamma+k_{||}v_{||}$.  A perpendicular gradient in electron momentum space, $\partial f/\partial p_\perp$, provides the dominant source of free energy to drive the maser. While the mechanism had been known for decades, e.g., \citet{Twiss1958}, a key insight by \citet{Wu1979} showed that when the relativistic correction is included in the resonance condition, many more electrons can be in resonance with the Doppler-shifted gyrofrequency. This is because, topologically, the resonance condition takes the form of an offset ellipse in the $v_\perp-v_{||}$ plane rather than a straight line ($\gamma=1$ case) or a circle centered on the origin of the distribution ($k_{||}=0$ case). As a result, far milder conditions are required for the CMI to be operative. 


In the case of planetary CMI emission, the gradients in $f({\bf p})$ required to drive the maser naturally occur for loss-cone, ring, and horseshoe distributions. A loss-cone distribution arises when electrons are trapped in a converging magnetic field configuration; e.g., a dipolar magnetosphere. Electrons with small pitch angles precipitate from the trap whereas those with larger pitch angles reflect and are trapped. Ring and horseshoe distributions result from convergent magnetic fields coupled with field-aligned electric fields \citep{Ergun2000}.  In the terrestrial case, the loss-cone instability is not the dominant driver of the CMI; rather, ring and/or horseshoe distributions drive the CMI \citep{Ergun2000}.  However, in the case of Jovian radio emissions, supported by recent observations by the {\sl Juno} mission, the loss-cone distribution is indeed relevant \citep{Louarn2017, Kurth2017}. We adopt Jupiter as our analog here and assume that the observed emission from UV~Ceti is the result of CMI emission from loss-cone electron distributions in a quasi-stable magnetosphere with energies of order 10 keV. We acknowledge that other particle distributions may prove to be relevant. 

The CMI may amplify electromagnetic waves in the extraordinary mode (x-mode) and/or the ordinary mode (o-mode). A key feature of the CMI is that when the electron plasma frequency $\omega_{pe}\ll \Omega_{Be}$, where $\omega_{pe}=(4\pi n_e e^2/m_e)^{1/2}$  and $n_e$ is the electron number density,  the x-mode is strongly favored for amplification just above the x-mode frequency cutoff $\omega_{x}={1\over 2}[\Omega_{Be}+(\Omega_{Be}^2+4\omega_{pe}^2)^{1/2}]\approx\Omega_{Be}(1+\omega_{pe}^2/\Omega_{Be}^2)$.  The amplified emission is beamed into a highly anisotropic pattern: the thin walls of a hollow cone.  For a shell distribution the emission is beamed perpendicular to the local magnetic field. For a loss-cone distribution the opening angle of the emission cone is determined by the resonance ellipse yielding maximum amplification. \citet{Hess2008} show that that the beaming angle can be expressed as $\theta_{bm}=\cos^{-1}[(v_{||\circ}/c)/(1- \Omega_{Be}/\Omega_{Be, max})^{1/2}]$ where $v_{||\circ}$ is the velocity on which the resonant ellipse is centered and $\Omega_{Be, max}$ is the electron gyrofrequency at the footpoint of the active magnetic field line.  The angular width of the walls of the emission cone is given as $\Delta\theta_{bm}< v_{||\circ}/c$ and the instantaneous bandwidth is $\Delta\omega/\omega \approx (v_{||\circ}/c)^2$.

\subsection{A Stellar Magnetosphere}


We explore the possibility that, like Jupiter, the radio emission from UV~Cet is the result of auroral CMI radio emission driven by a loss-cone anisotropy in a dipole-like magnetosphere. We adopt a dipole field strength at the pole of $B_p=2000$ G (O. Kochukhov, private communication) and a magnetic moment $M=B_pM_\star^3/2$, where $M_\star$ is the stellar radius. Our schematic model is similar in some respects to those used to model Jovian auroral arcs of DAM radio emission in dynamic spectra; e.g., \citet{Hess2008, Hess2011, Louis2019}. The rotational axis has an inclination $i\sim 60^\circ$ and the star is assumed to rotate with a period $P_{rot}=5.45$ hrs. We take the $z'$-axis to aligned with the rotation axis, the $x'$-axis such that the observer line-of-site is in the $x'-z'$ plane, and the $y'$-axis completes the right-handed frame. In the frame of the magnetosphere, spherical coordinates are convenient. Ignoring the inclination, we have the co-latitude $\theta$ measured relative to $z'$, the longitude $\phi$, and radius $r$.  We take $\phi=0$ to be in the $x'-z'$ plane. The coordinates are related through 

\begin{equation}
x'=r\sin\theta\cos\phi;\qquad y'=r\sin\theta\sin\phi;\qquad z'=r\cos\theta
\end{equation}

\noindent The corresponding components of the magnetic field vector for a dipolar field are for the stellar magnetic moment $M$

\begin{equation}
B_x'={{3Mx'z'}\over r^5}; \qquad B_y'={{3My'z'}\over r^5}:\qquad B_z'=M{{(3z'^2-r^2)}\over r^5}
\end{equation}

\noindent A rotational transformation then corrects for the inclination of the rotation axis from $(x',y',z')$ to $(x,y,z)$ where the x-axis is now directed along the line of sight and the z-axis is tilted $90-i$ degrees from the rotational axis. It is convenient to discuss specific magnetic field lines. A useful parameter that relates the radius $r$ to the co-latitude $\theta$ along a dipolar magnetic field line is $L$, defined through $r=L\sin^2\theta$. When $\theta=90^\circ$, $r=L$. 
We have indicated regions along a sample of magnetic field lines were $\Omega_{Be}$ falls within the observed frequency range of 886-1682 MHz. The corresponding radial ranges depend on $L$. For example, for $L=2$ the radial range is 1.34-1.52 $R_\star$ and for $L=5$ it is 1.48-1.53 $R_\star$. It is seen that resonant frequencies occur at conjugate co-latitudes over the northern and southern hemispheres along any given field line. 

The CMI results in emission from a specific location along a given field line into a beaming pattern that is at a large angle to the local magnetic field vector $\theta_{bm}$, given above.  If $(b_x,b_y,b_z)=(B_x/|B|, B_y/|B|, B_z/|B|)$ the magnetic field orientation at that location is $(\theta_\circ,\phi_\circ)=[\sin^{-1} b_z, \tan^{-1}(b_y/b_x)]$ and the angle of the  field line relative to the line of sight angle is $\theta_{los}=\cos^{-1}(\cos\theta_\circ\cos\phi_\circ)$. The emission is beamed toward the observer when $|\theta_{bm}-\theta_{los}|<\Delta\theta_{bm}$. A given source location beaming radio emission toward the observer lies on a particular field line that we refer to as an ``active field line" (AFL). An AFL is one to which fast electrons have access and can produce a loss-cone distribution that is unstable to CMI emission. 

\begin{figure}
\begin{center}
\plottwo{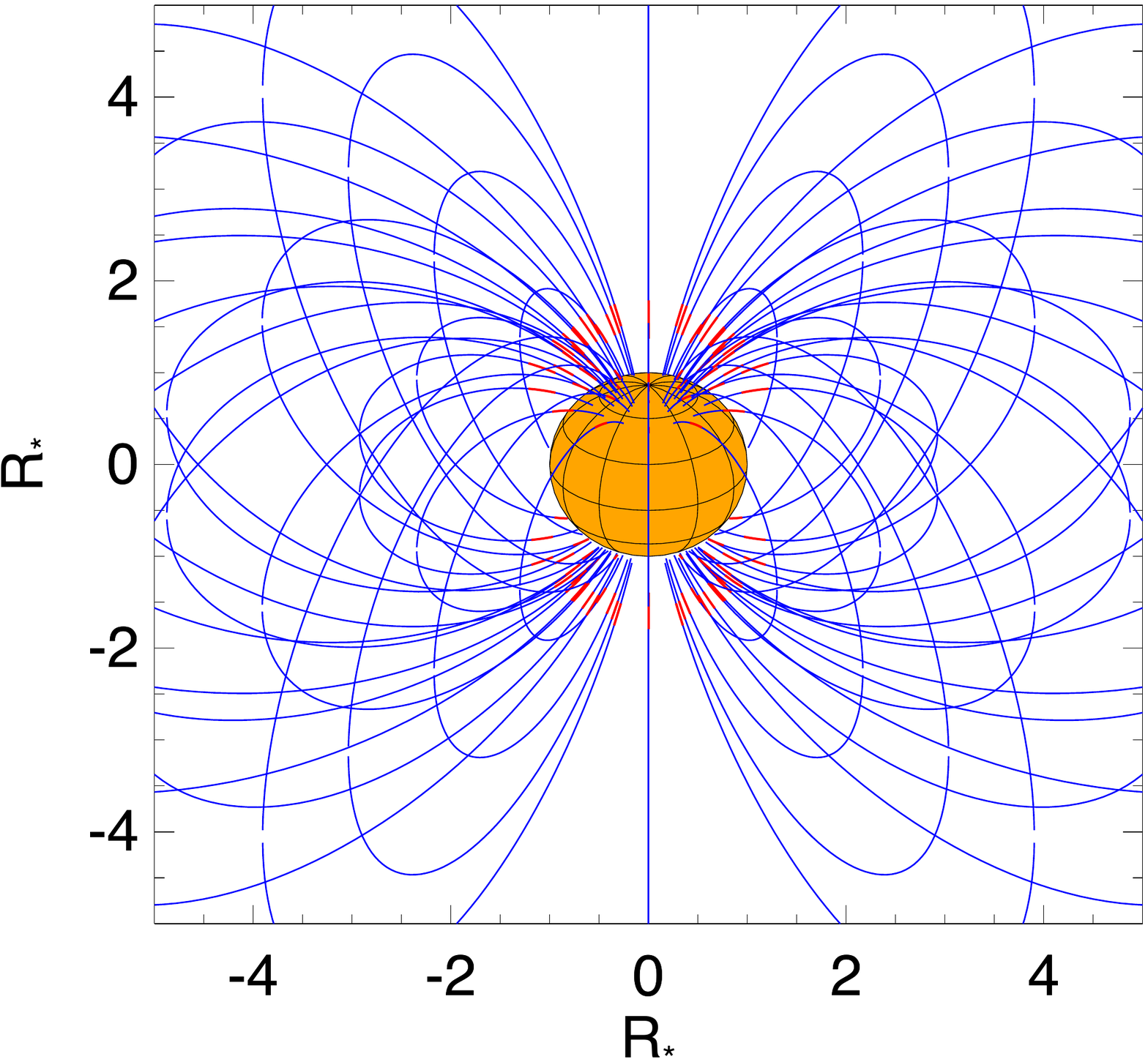}{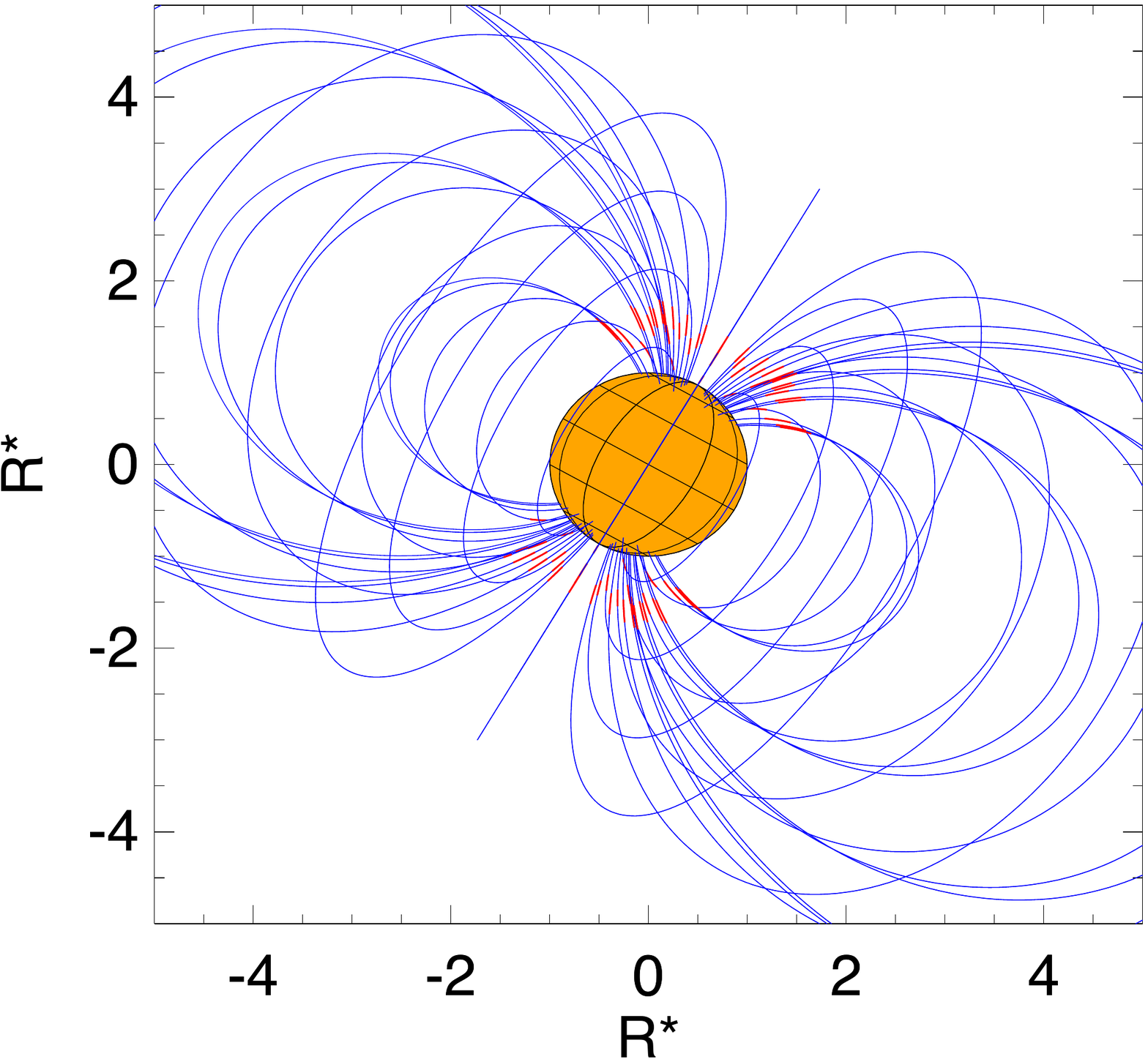}
\caption{Two perspectives of the stellar magnetosphere. {\sl Left}: as seen by the observer; {\sl Right}: as seen from the side. The direction of the observer is to the right. Dipolar field lines are plotted for $L=3,5,7,$ and 9. The red highlights on each field line indicate where the cyclic electron gyrofrequency $\nu_{Be}$ falls with in the observed frequency band of $886-1682$ MHz. }
\end{center}
\end{figure}

\subsection{The Plasma Environment}

We now consider the plasma environment around UV~Cet. UV~Cet is a strong X-ray emitter with a quiescent luminosity of a few $\times 10^{27}$ ergs-s$^{-1}$ in the 0.1-10 keV band, comparable to that of the Sun; flares can exceed this base X-ray luminosity by two orders of magnitude \citep{Audard2003}. The hot plasma producing the X-ray emission has temperatures $3-6\times 10^6$ K and a volume emission measure $EM\gtrsim 10^{49}$ cm$^{-3}$. The details of how the hot plasma is distributed around the star are not well constrained observationally.  Given that $\langle B\rangle \sim 4$ kG or more at the photospheric level a substantial fraction of the plasma may be confined to closed magnetic structures near the star. On the other hand, in light of the the large scale dipolar field and rapid rotation of UV~Cet it is worth considering whether the magnetosphere conforms to a rigidly rotating magnetosphere \citep{Townsend2005, UdDoula2008}. Here, the plasma distribution is determined in part by gravitational and centrifugal accelerations. \citet{Townsend2005} define the Kepler co-rotation radius $r_k=(GM_\star/\Omega^2)^{1/3}$ -- $G$ is the gravitational constant, $M_\star=0.12$ $M_\odot$ is the stellar mass, and $\Omega=2\pi/P_{rot}$ is the star's angular rotation rate -- where the gravitational and centrifugal forces on a parcel of plasma balance. For radii such that $r< r_K$ the joint potential is dominated by gravity and is essentially spherical, but for $r> r_K$ it is cylindrical with a minimum at the equator for an axisymmetric dipole. As a result, plasma accumulates as an equatorial disk that is truncated at an inner radius $r_i\sim 0.87 r_K$ and an outer radius of perhaps $r_o\sim 2 r_K$. For UV~Cet, we have $r_i\approx 4$ R$_\star$ and $r_o\sim 10$ $R_\star$, respectively. Unlike a Keplerian disk, which flares upward on either side of the equator, Townsend \& Owocki argue that the density distribution perpendicular to the disk is well described by a Gaussian stratification with a near-constant width $h_m\sim(2kT/3m_p\Omega^2)^{1/2}$ which, for UV~Cet is $h_m \sim$ a few $\times 0.1T_6^{1/2}$ r$_K$ where $T_6$ is the temperature in units of $10^6$ K. Hence, depending on the temperature of the equatorial the disk may be thin or relatively thick. The maximum mass density in the equatorial disk is determined by equating the kinetic energy of the co-rotating plasma with the magnetic energy density \citep{Linsky1992}, resulting in a maximum electron number density of $n_e\sim (B_e^2 P_\star^2/16\pi^3R_\star^2 m_p)/L^8 = 3.8 \times 10^{15}/L^8$ cm$^{-3}$. The maximum density therefore ranges from $6\times 10^{10}$  cm$^{-3}$ at $r_i$ to $4\times 10^7$  cm$^{-3}$ at $r_o$.

What is the source of the plasma? For $r<r_k$, eruptive flares and coronal mass ejections (CMEs) may contribute a significant amount of hot plasma but it must ultimately fall back to the star. Given the strong and large scale magnetic fields on UV~Cet, CMEs may in fact be suppressed \citep{AlvaradoGomez2018}. For $r_i<r<r_o$ mass accumulates in an equatorial disk as the result of a tenuous wind. For $r\gtrsim 2r_K$ (magnetically linked to polar regions with colatitude $\lesssim 18^\circ$) mass loss may be dominated by wind since the fraction stellar surface potentially contributing flares is only 10\%.  Mass loss due to a wind is poorly known for late-type dwarfs. For the M3.5 dwarf EV~Lac \citet{Wood2005} estimate a mass loss rate comparable to that of the Sun while that of the M5.5 dwarf Prox Cen is estimated to be $<0.2$ times solar. We note that if the mass loss rate per unit area on UV~Cet is comparable to that of the Sun, in which case the mass loss rate would be $\sim 10^{-15}$ M$_\odot$-yr$^{-1}$. As we show in \S4.5.2, however, the mass loss rate may be significantly smaller than this. We conclude that hot plasma is likely confined in closed magnetic structures close to the star with a more extended component possibly distributed in the corona and equatorial disk. 

\citet{Benz1998} detected and mapped radio emission from UV~Cet at 8.4 GHz using VLBI techniques. They found a double source that they interpreted to be nonthermal gyrosynchrotron emission from sources associated with the magnetic poles of UV~Cet, each at a height of $\sim\!2$ R$_\star$. The electron energies required to drive CMI emission are modest -- 10s of keV -- whereas those needed to produce nonthermal gyrosynchrotron emission are much greater: several MeV. Hence, UV~Cet must not only produce and maintain substantial SXR-emitting thermal plasma, it must produce and sustain a source of the fast electrons needed to drive coherent CMI emission, as well as the much more energetic electrons needed to produce the incoherent nonthermal gyrosynchrotron emission. 


\subsection{Spectral Signature of CMI Emission}

The radio outburst from UV~Cet is highly RCP. Coupled with the inclination of the star and positive polarity of the polar magnetic field \citep{Kochukhov2017a}, the emission is therefore predominantly in the sense of the x-mode. There is no evidence of harmonic structure in the dynamic spectrum. We therefore assume the observed radiation is CMI emission near the fundamental ($s=1$) of the electron gyrofrequency. The CMI drives fundamental emission in the x-mode when $\omega_{pe}/\Omega_{Be}\ll 1$. While it is likely that a range of electron energies are unstable to the CMI mechanism, for simplicity we assume moderately fast electrons are responsible for the emission, with $v=0.3c$, corresponding to 25 keV. The opening angle of the emission pattern is then $73^\circ$ relative to the local magnetic field vector. 

The observed dynamic spectrum of the radio outburst from UV~Cet is complex, showing a variety of structures that drift with frequency in time. We first consider whether the observed drifts can be understood in terms of discrete sources that i) meet the conditions necessary for the CMI mechanism to be operative; ii) are in a location that will beam the resulting emission toward the observer. The spectral signature will then evolve in time due to the rotation of the star alone, as opposed to a moving exciter that is unrelated to stellar rotation. As shown in Fig.~7, a range of radii in both the northern and southern polar regions satisfy the frequency resonance condition for the range of frequencies sampled by MeerKAT (886-1682~MHz). Since the opening angle of the emission cone is at a large angle relative to the magnetic field vector in this schematic model ($\theta_{bm}\approx 73^\circ$) only a restricted range of longitudes toward either limb will beam toward the observer - the details depend on the $L$ value of the AFL and the hemisphere in which the source occurs as we now discuss in greater detail. 


Consider a location on an AFL with $L=r\sin^2\theta$ at some longitude $\phi$ where conditions (i) and (ii) are met at a frequency that falls within the MeerKAT bandpass.  Even if the source persists, stellar rotation will move the emission pattern (the walls of an open cone) until it is no longer beamed toward the observer. The source will therefore appear in a dynamic spectrum as a pulse of emission at a particular time and frequency with some bandwidth $\Delta\nu/\nu\lesssim 100$ MHz for $v/c=0.3$. The duration will be no longer than the time required for the cone wall to sweep past the observer. With $\Delta t\sim 10-60$s the angular width of the cone wall is $\Delta\theta_{bm}\sim 0.2-1^\circ$. If the source meets conditions for the CMI instability over a range of distance along an active field line a range of frequencies will be actively in emission. However, an AFL will not appear to emit at the full range of resonant frequencies at the same time because of the angle of the source relative to the magnetic field varies as a function of location along the field line and, hence, with frequency. Instead, a source at a lower frequency may lead or lag a source at a higher frequency, depending on whether the active field line is rotating toward the observer or away from the observer. This is illustrated in Fig.~8, where we show the locations along an AFL of different frequencies when the longitude of the AFL is such that the CMI emission is beamed toward the observer. That is, each line corresponds to a specific time in this rendering. For this example we have $L=2$, frequencies at 100 MHz intervals from 900 to 1600 MHz, and electrons with $v/c=0.3$. The stellar longitudes at which CMI radiation is beamed toward the observer range from $-64^\circ$ at 1600~MHz to $-54^\circ$ at 900~MHz. In this example, therefore, the AFL emits toward the observer at 1600 MHz approximately 9 min before it emits toward the observer at 900~MHz.

\begin{figure}
\begin{center}
\plottwo{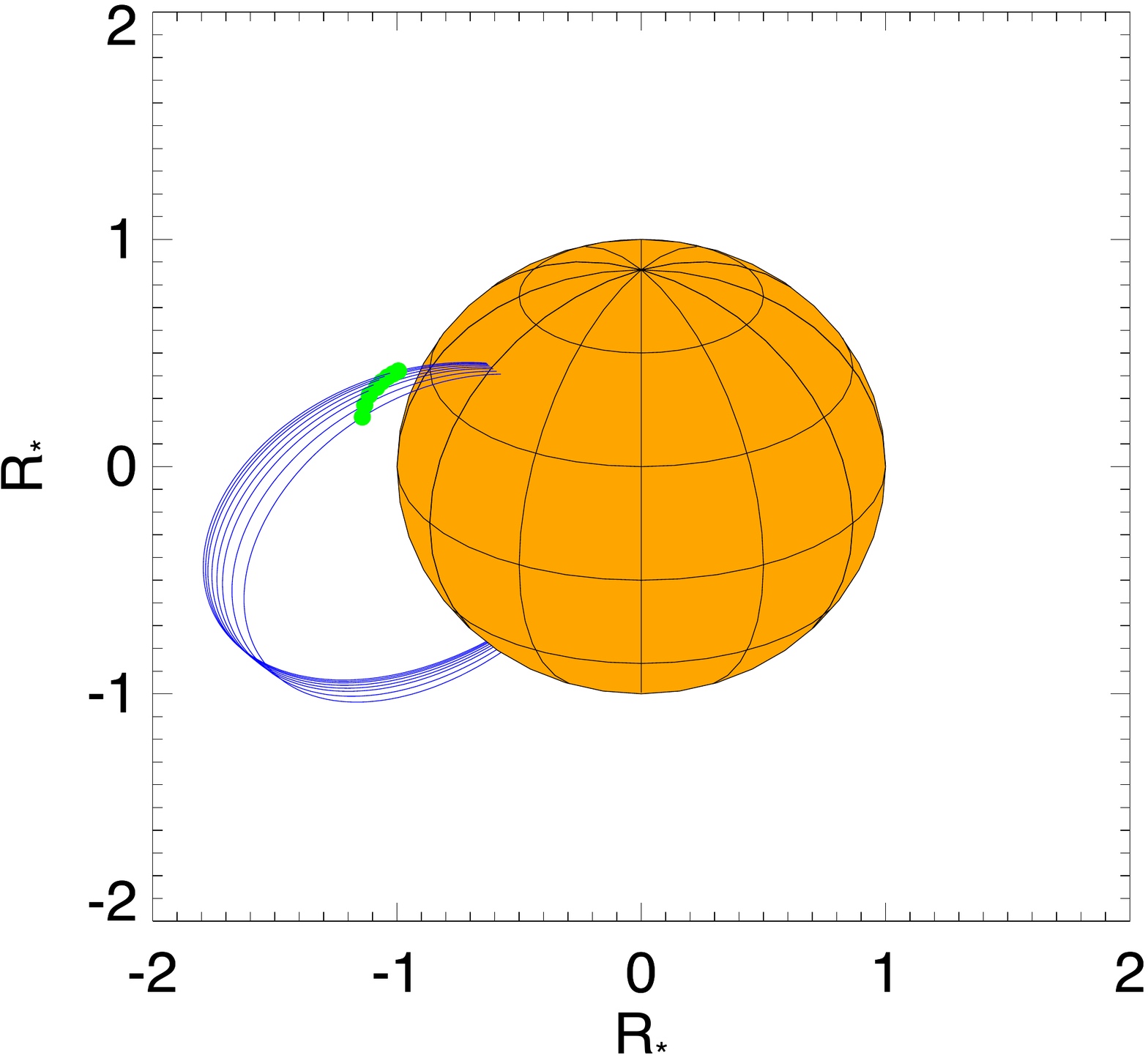}{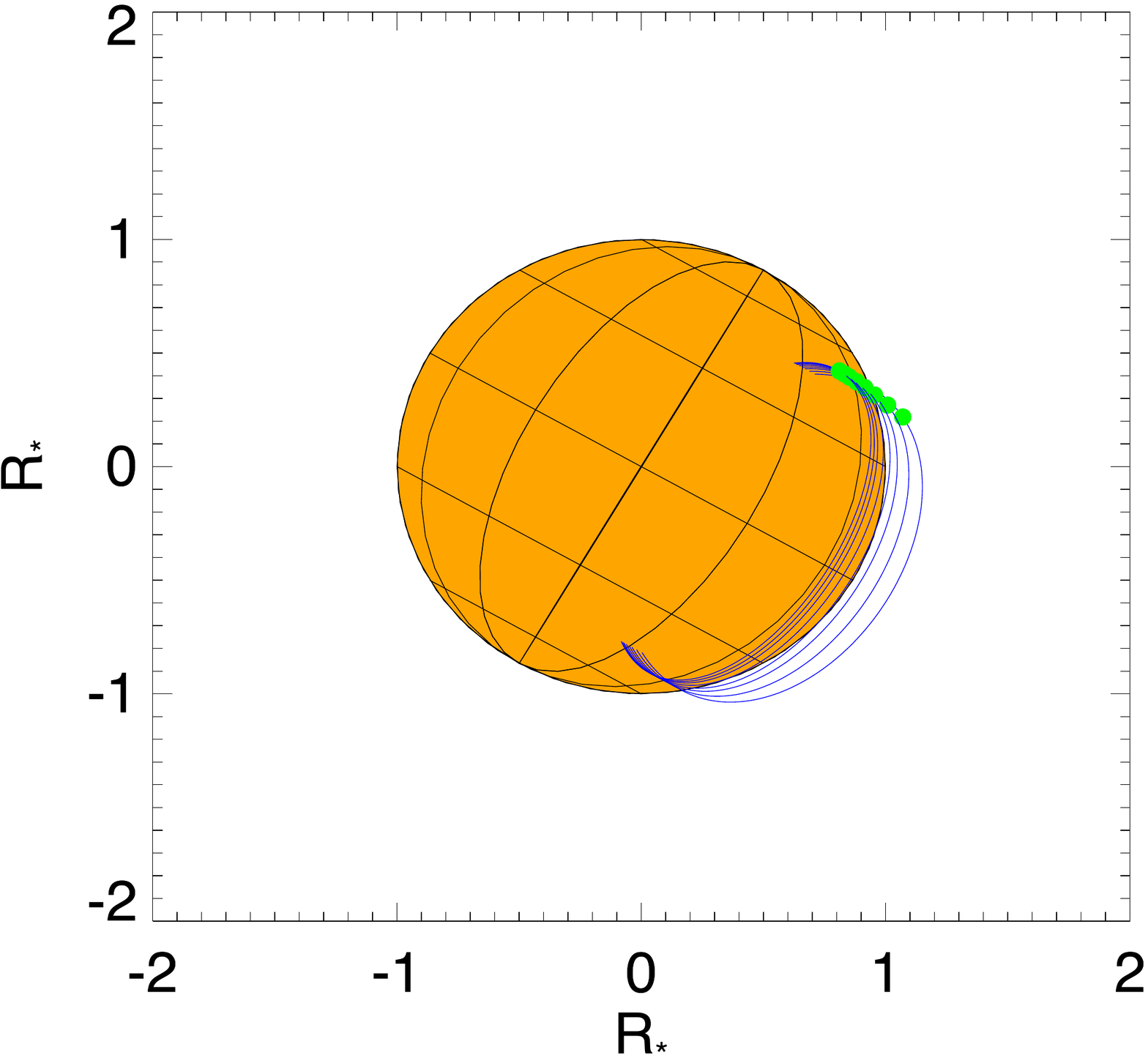}
\caption{Two perspectives of an active field line at $L=2$. {\sl Left}: as seen by the observer; {\sl Right}: as seen from the side where the direction of the observer is to the right. Dipolar field lines are plotted for successive times and longitudes for when the source on the active field line is beaming CMI radiation toward the observer. The green highlights on each field line indicate the location of the source at the time it is beaming toward the observer at frequencies 900 to 1600~MHz at 100~MHz intervals. The corresponding longitudes range from $-54^\circ$ to $-64^\circ$. The higher frequency emission leads the lower frequency emission in this case. }
\end{center}
\end{figure}

We have ``fit" several features in the dynamic spectrum to emission expected from an AFL rotating toward $(F_1,F_2,F_3)$ and away from the observer $(F_4,F_5)$, shown as contours in the lower panel of Fig. 5. The values of $L$ for the AFLs corresponding to features $F_1$ through $F_5$ are $[4.7,2.15,1.8,1.7,1.5]$, respectively, while the ranges of longitude are $[(-85^\circ,-84^\circ), (-67^\circ,-59^\circ), (-58^\circ,-39^\circ), (+48^\circ,+54^\circ), (+16^\circ,+39^\circ)]$. Our assumption, within the confines of this simple model, is that features in the outburst - peaks A, B, C -  involve multiple AFLs distributed over a range of longitudes moving toward or away from the observer due to stellar rotation. Consider peak A. It displays multitudes of short-duration bursts of limited bandwidth within an envelop of AFLs distributed over a longitudinal range $\Delta\phi_A\sim 17^\circ$ at $L\approx 4.7$ that is moving toward the observer. Sources along each AFL emit intermittently. In similar fashion, the first half of peak B contains structures that can be fit with sources located on AFLs with $L\approx 2$ rotating toward the observer. The range of longitudes needed to account for the structures drifting from high to low frequencies in peak B is $\Delta\phi_{B1} \approx 25^\circ$. It is difficult, however, to reconcile peak B with a single range of AFL longitudes where they are first approaching the observer and then receding to account for the low-to-high frequency drifts seen toward the end of peak B. Instead, a separate set of AFLs must account for the emission with $\Delta\phi_{B2}\approx 10^\circ$. That is, an advancing range of AFLs and a receding range overlap in time in the spectrum as peak B but are  separated by $\sim 90^\circ$ in longitude - at least within the context of this simple model. While some structures in peak C are consistent with receding AFL emission at $L\sim 4$, the slow-drift striations require receding AFLs with $L<2$. 

There are also source locations in the southern hemisphere that meet conditions (i) and (ii). We considered $2<L<10$ and find that such sources can be found for $L>2$ but in all cases the source longitudes require $|\phi|>83^{\circ}$. However, the stellar inclination is such that radiation from a southern auroral source at these longitudes must propagate along a path for which the magnetic field strength first increases before decreasing. The emission is therefore encounters a stop band at the x-mode cutoff frequency and does not propagate. Within the context of this simple model, we only expect to see CMI emission from sources emitting from restricted ranges of longitude in the northern hemisphere; i.e., RCP only. 

We make no attempt here to produce a fully self-consistent arrangement of longitudinal bands of AFLs that are in agreement with the observations. The present attempt is more along the lines of a plausibility argument. Following the ideas introduced above, a given feature A, B, or C in the radio outburst corresponds to a longitudinal band of AFLs, each of which emits intermittently in time and frequency. Fast electrons are presumably streaming down along a given AFL. Some of these precipitate into the low atmosphere and the rest mirror, setting up an upward loss-cone distribution function that is unstable to the CMI. The precipitating electrons are stopped by the dense, cool atmosphere of UV~Cet where they may emit O/UV/SXR radiation as auroral ovals or partial ovals. We suggest that perhaps some of the low-level optical and SXR variations observed \citep{Fleming2022, Audard2003} are the result of precipitating non-thermal electrons with energies of 10s of keV. An important question is where the fast electrons originate and how they are accelerated. We can only speculate. One possibility is that interactions in an equatorial plasma disk, or in a breakout current sheet, result in fast electrons, an idea explored by in the context of chemically peculiar stars \citep{Linsky1992, Townsend2005, UdDoula2008}, although \citet{Palumbo2022} recently argued that centrifugal mass breakout is relevant to an M3.5 dwarf. While this idea may be plausible for aspects of peaks A and C, for which AFLs at $L\sim 4$ are consistent in location with the inner radius $r_i$ of an equatorial plasma disk, it is hard to reconcile with the strong emission seen in peak B, which requires  smaller values of $L\sim 2$ for the relevant AFLs. The evidence for sources on AFLs with $L\sim 10$, which would map into the outer part of an equatorial disk where centrifugal mass breakout might occur, is weak. We see no obvious way out of this puzzle but note that while the large scale field of UV~Cet is dominated by an axisymmetric dipole, multipolar components become increasingly important close to the star. Since the resonant radii are only $\sim 1.3-1.5$ R$_\star$ higher order components likely result in departures from a strictly dipolar field, perhaps producing analogs to solar coronal  streamers and pseudo-streamers.

We end this section with a brief discussion of how CMI fundamental x-mode radiation escapes from the source to the observer. As pointed out by \citet{Melrose1982} thermal gyroresonance absorption of CMI radiation in resonant layers overlying the source at harmonics $s\ge 2$ can be catastrophic within the context of a stellar corona. The optical depth of a given resonance layer to gyroresonance absorption is given approximately by

\begin{equation}
\tau_s \approx {\omega_{pe}^2 \over \omega^2} {{s^{2s} 2^{1/2}}\over {2^{s-1} s!}} {{\omega L_B}\over c}\Bigl({{k_BT_e}\over {m_e c^2}}\Bigr)^{s-1} 
\end{equation}  

\noindent The absorption is strongest at the second harmonic layer $s=2$. Previous workers, assuming parameters suitable for solar flares, found $\tau_2$ of a few $\times 10^3$! Since the optical depth has a dependence of $\sin^2\alpha(1+\cos\alpha)^2$ for x-mode, it has been suggested that if the radiation were scattered to small angles - in an under-dense cavity, for example - it could escape \citep{Ergun2000}.  This may not be necessary. Assuming a magnetic scale height $L_B=R_\star/3$, $T_6=1$ MK, and $\omega/2\pi=1$ GHz the optical depth at the second harmonic layer will be $\tau_2\sim 10^{-6} n_e T_6$. For coherent emission a degree of absorption is acceptable. \citet{Zic2019} estimated the lower limit to the brightness temperature of the radio outburst to be $0.22-4.3\times 10^{13}$ K. The intrinsic brightness temperature of fundamental x-mode radiation can be considerably higher. For example, brightness temperatures of Jovian CMI emissions generally exceed $10^{15}$ K \citep{Zarka1998}.  For $\tau_2 \sim 3$ the brightness is attenuated by a factor of 20 and the electron number density can be as high as $10^6$ cm$^{-3}$ in the harmonic layer for X-ray emitting temperatures and perhaps $5\times 10^6$ for somewhat cooler plasma temperatures.  Interestingly, these values are similar to those of the solar wind at $\sim 2$ R$_\odot$ and would be similar at the corresponding radius for UV~Cet if its mass loss rate per unit area is similar to that of the Sun. These should be regarded as upper limits, however. As we show in \S4.5.2 the electron number density may be less than estimated here, further alleviating the difficulty of CMI radiation escaping the source. These estimates are also consistent with assumptions made regarding small value of $\omega_{pe}/\Omega_{Be}$ in the CMI source. Since the harmonic layer is at a radius that is a factor $\sim\!2^{1/3}$ greater than the source radius, the density in the source will be a factor of a few higher than in the harmonic layer. The CMI has maximum growth in the x-mode when $\omega_{pe}/\Omega_{Be}\ll 1$. For plasma densities of $3\times 10^6$ to $1.5\times 10^7$ cm$^{-3}$ and $\Omega_{Be}/2\pi=1$ GHz, we have $\omega_{pe}/\Omega_{Be}\approx 0.016-0.035$. We conclude that conditions in the CMI source are consistent with those required for the amplification of fundamental x-mode radiation and that it can escape to the observer without being fully absorbed at the harmonic layer. 

\subsection{Polarization}

We now consider the polarization of the observed emission. We first consider the circularly polarized emission and then the surprising presence of elliptically polarized components.

\subsubsection{Circular Polarization}

The observation of strong RCP emission is consistent with other observations of UV~Cet near a frequency of 1~GHz, both recent \citep{Villadsen2019, Zic2019} and those that are several decades old \citep{Kundu1988}. The high fractional degree of circularly polarized emission in the sense of the x-mode is fully consistent with the assumption that the operative emission mechanism is the CMI, as noted above and elsewhere \citep{Zic2019}. The emission is not 100\% circularly polarized, however, as might be expected if the emission were due solely to fundamental x-mode. The degree of polarization ranges from 0.1-0.75 RCP when Stokes I exceeds 5~mJy. Among possible reasons for a fractional circular polarization less than unity are: that there is a separate source of radiation that is predominanty LCP; that the RCP emission is depolarized as it propagates to the observer; or some combination of the two. 
It is possible, even likely, that the conjugate footpoint (southern hemisphere) of an AFL is also emitting fundamental x-mode radiation. Since the polarity of the local magnetic field is opposite that in the northern hemisphere, the emission would be LCP. However, for reasons given in \S4.4, such emission is not expected to escape to the observer. The alternative is that a depolarization mechanism is operative on the CMI emission.

\begin{figure}
\begin{center}
\includegraphics[width=5in]{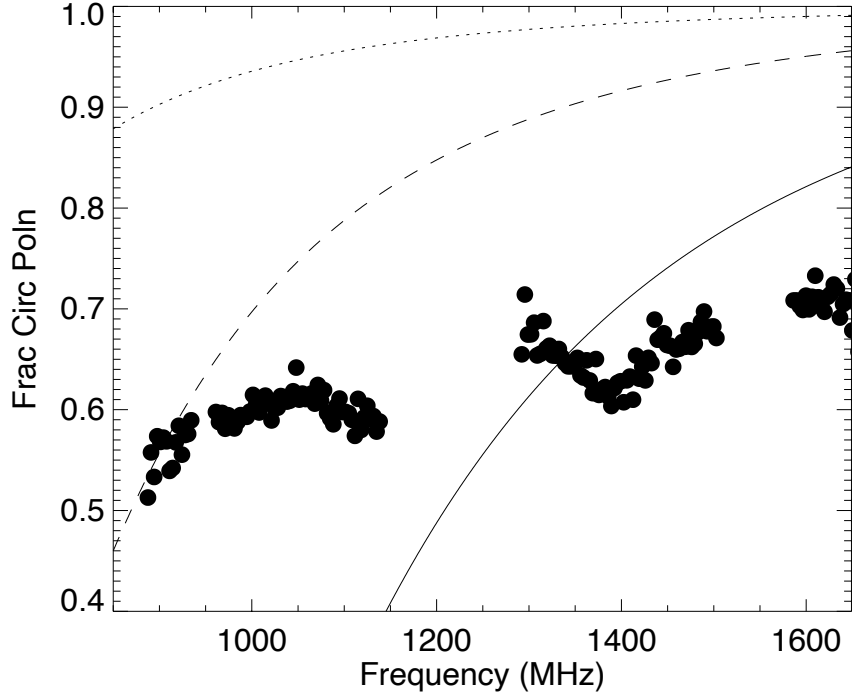}
\caption{The fractional degree of circular polarization across the MeerKAT frequency band for a circularly polarized wave incident on a QT region. The magnetic field is taken to be 130, 100, and 70 G, and the electron number density is $1.7 \times 10^6, 10^6$, and $5.8 \times 10^5$ cm$^{-3}$, for the solid, dashed, and dotted lined, respectively. The filled circles are the measured fractional polarization near the peak of peak B.}
\end{center}
\end{figure}

A promising candidate is magnetoionic mode coupling (see e.g., \cite{Budden2009} or \cite{Melrose1980} for formal developments of the theory) under conditions of ``quasi-transverse" (QT) propagation. \citet{Cohen1960} defined a coupling ratio $Q$ to demarcate strong ($Q\ll 1$) and weak ($Q \gg 1$) mode coupling and showed that for so-called ``quasi-longitudinal" (QL) propagation, the x- and o-modes were weakly coupled at radio wavelengths. For QT propagation, however, where the radiation traverses a region where the angle between ${\bf k}$ and ${\bf B}$ changes from $<\pi/2$ to $>\pi/2$, conditions of strong mode coupling may be achieved. The details of mode coupling in a QT region are complex and will not be reviewed here. In brief, x-mode radiation can couple into o-mode to produce a linearly polarized component as it traverses a QT region. The linearly polarized component is subsequently eliminated by differential Faraday rotation, leaving partially circularly polarized emission. \citet{Zheleznyakov1964} developed a theory of depolarization on QT regions in the solar corona and showed that if a 100\% circularly polarized mode is incident on a QT region with a coupling ratio $Q$ the fractional polarization of the emerging wave is $m_{circ}=-(1-2e^{-x})$, where $x=\pi/2Q$ and

\begin{equation}
Q=\Biggl({\omega\over {4c}} {{X Y^3}\over{\theta'}}\Biggr)^{-1}
\end{equation}

\noindent where $X=\omega_{pe}^2/\omega^2$ and $Y=\Omega_{Be}/\omega$. Clearly, if $Q\ll 1$ (weak coupling), the incident radiation simply reverses its sense of circular polarization upon encountering a QT region but remains in the same mode, but if $Q\gg 1$ (strong coupling) the sense of polarization remains the same. For intermediate values the fractional polarization is reduced. Hence, if conditions in the magnetospheric environment of UV~Cet meet those required for QT propagation, depolarization may occur. In the source we have $\omega=\Omega_{Be}$ and so $X\ll1$ and $Y=1$. For $\theta'\sim 1/R_\star$, $Q\ll 1$ in the source and depolarization does not occur. However, somewhat farther from the source it may be possible that QT regions are encountered that act to partially depolarize the emission. Fig.~9 shows the fractional circular polarization across the MeerKAT frequency band to emerge from a QT region for plausible plasma parameters for radii ranging 2-2.4 $R_\star$. An obvious consequence of this simple picture is that the degree of depolarization is larger at lower frequencies. We therefore also plot an example of the $m_{circ}$, the degree of circular polarization at 22:00, near the maximum of peak B. While expectations are qualitatively met, the observed fractional polarization as a function of frequency is flatter than might be expected. Note, however, that no provision has been made for the fact that at a given time, the emission at each frequency is from a different AFL and height. One might also suppose that for a given source of CMI emission subject to depolarization, a constant fraction of RCP emission would be converted to LCP emission. This is unlikely to be the case because plasma conditions between the source and the observer change continuously due to stellar rotation. Indeed, it may be possible to leverage the details of the RCP and LCP emission as a function of frequency and time to constrain the nature of the depolarizing medium. The viability of depolarization on QT regions and its possible use as a diagnostic of the plasma medium needs more detailed modeling than the simple considerations offered here. 

It is also interesting that there are times when the emission is weakly LCP; specifically, during brief times before, and in between, peaks A and B (Fig.~4). During these times the total intensity is low -- 5-10 mJy -- but above the quiescent flux density of 4.2 mJy. We speculate that it may be that during these brief times when CMI emission ceases, underlying incoherent nonthermal gyrosynchrotron emission associated with the outburst is seen. Optically thick gyrosynchrotron emission can be o-mode polarized for sources in a well-ordered magnetic field but any source of optically gyrosynchrotron emission would necessarily need to be from a different location than the CMI source. This idea, too, requires further investigation. 

\subsubsection{Elliptical Polarization}

Elliptically polarized emission from radio outbursts near 1~GHz on UV~Cet was first reported by \citet{Zic2019} and is confirmed here. This is a remarkable finding because significantly elliptically polarized radio sources are believed to be quite rare in nature: it was previously only thought to occur in certain decameter (DAM) radio bursts on Jupiter and on some pulsars. \citet{Melrose1991} considered the case of DAM bursts from Jupiter in the context of the loss-cone-driven CMI in a dipolar magnetic field. They suggest that the bursts are intrinsically elliptically polarized and employ an argument based in strong mode coupling to conclude that it was possible to produce such radiation provided the plasma density in and around the source is very low: $\lesssim 5$ cm$^{-3}$ in the Jovian case. 
In strict analogy to the Jovian case, \citet{Zic2019} conclude that, scaling up to a frequency of 1~GHz (using Eqn.~14 of \citet{Melrose1991}), the implied plasma density in and around around the CMI source on UV~Cet is likewise very low, $n_e\lesssim 40$ cm$^{-3}$. Note, however, that this estimate should be reduced by a factor $R_J/R_\star\approx 1.6$ to an even lower density limit. It is worth asking, however,  whether the Jovian analog applies. The answer depends on the plasma environment surrounding the source. We found above that the CMI instability is operative and the emission can escape in the presence of plasma densities that are many orders of magnitude greater than is necessary if the elliptical polarization is intrinsic to the source. However, if the linearly polarized fraction is not ``born" it must be ``made"; that is, the linearly polarized component to the radiation is acquired as it propagates away from the source. 

\begin{figure}
\begin{center}
\includegraphics[angle=90,width=7in]{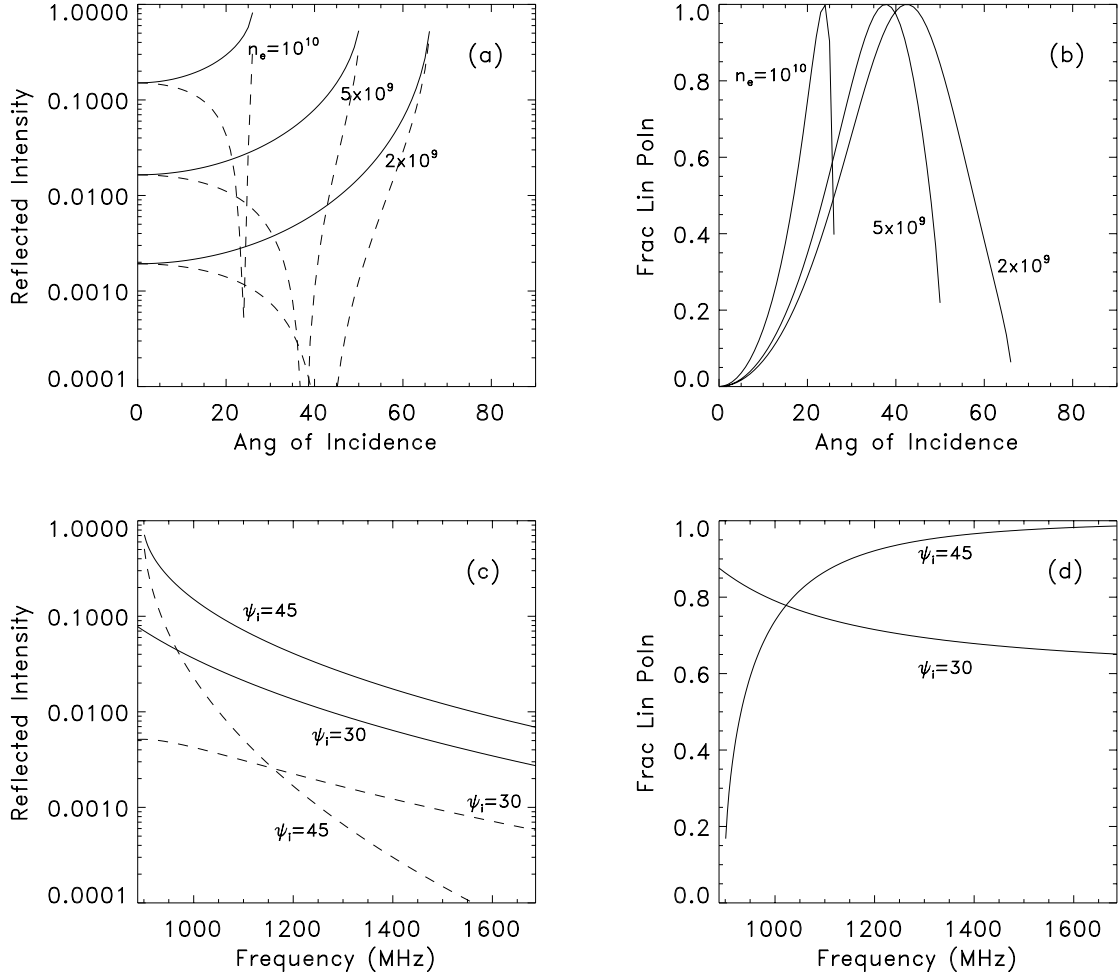}
\caption{a) Intensity of the reflected radiation for when the electric field vector is perpendicular (solid line) and parallel (dashed line) to the plane of incidence. The electron number density in the slab is indicated in each case. b) The fractional linear polarization in each case. Panels c, d) The same as panel (a) and (b), but as a function of frequency for two incident angles and for a plasma density of $5\times 10^9$ cm$^{-3}$. }
\end{center}
\end{figure}

One possibility is that the RCP waves reflect from an over-dense plasma structure some distance away from the source -- perhaps as much as several stellar radii.  The general problem of the reflection of electromagnetic waves from a plasma at oblique incidence requires a kinetic treatment \citep{Angus2010}. However, for the simplified case of oblique incidence on a cold, collisionless, plasma slab where $\omega\gg \Omega_{Be}$ the reflection coefficients reduce to the usual Fresnel equations at a dielectric interface. The refractive index of the medium for the incident wave is $\mu_i\approx 1$ and an over-dense plasma structure will have a refractive index $\mu_o=(1-X)^{1/2}$. A purely circularly polarized wave is a combination of orthogonal linearly polarized waves with a relative phase difference of $\pi/4$. If we take the two components to be such that the electric field vector is perpendicular (s-polarized) and parallel (p-polarized) to the plane of incidence the reflection coefficients are 

\begin{equation}
R_{\perp} = \Biggl| {{\cos\psi_i-\mu_o\cos\psi_o}\over{\cos\psi_i+\mu_o\cos\psi_o}}\biggl|^2 \qquad
R_{||} = \Biggl| {{\cos\psi_o-\mu_o\cos\psi_i}\over{\cos\psi_o+\mu_o\cos\psi_i}}\Biggr|^2 
\end{equation}

\noindent where $\psi_i$ is the angle of incidence, $\psi_o$ is the angle of transmission into the plasma slab, and the two are related via Snell's law as $\sin\psi_i=\mu_o\sin\psi_o$. Reflection of the RCP CMI emission from the over-dense structure at an incident angle equal to Brewster's angle $\theta_B=\tan^{-1}\mu_o$ results in $R_{||}=0$, the result being a reflected wave that is 100\% linearly polarized in the sense perpendicular to the plane of incidence (s-polarized). In fact, the reflected radiation can be significantly linearly polarized for a range of angles, as shown in Fig.~10 for a frequency of 1~GHz for several plasma densities. Reflection is most effective as the frequency of the incident radiation approaches the plasma frequency -- in the sense of maximizing both the fractional linear polarization and reflected intensity. Also shown is the reflected intensity and the fractional linear polarization as a function of frequency for two incident angles and a plasma density of $5\times 10^9$ cm$^{-3}$. In both examples (Fig.~10c) the intensity of the polarized emission decrease with frequency, as noted for bursts $l_2$ and $l_3$. The degree of polarization, however, declines at low frequencies in one case ($\psi_i=45^\circ$) but is relatively constant for the other case ($\psi_i=30^\circ$), as observed. It appears possible to accommodate the observations. Note that for $\psi_i>\sin^{-1}\mu_o$ the incident radiation is completely reflected (total internal reflection) and no linearly polarized component is expected under these condition although it may offer another avenue for CMI radiation to enter into the line of sight.  A possible point in support of this scenario is the diminution in Stokes I and Stokes V intensity at the time of peak $l_1$ noted in \S3.2 and Fig.~6 since it is expected that a fraction of the incident intensity will not propagate. 

A requirement for this scenario is that the differential Faraday rotation between the location of a reflection and the observer, for which conditions of QL propagation are assumed to apply, is small enough that the linearly polarized component persists. Although the SNR is poor, the EVPA does not seem to change by a large amount over the roughly 900-1100 MHz frequency range in which the linearly polarized components are observed. Faraday rotation is given by $\chi={\rm RM}\lambda^2$, where RM is the rotation measure. We also have $d\chi/d\nu=-(2c^2/\nu^3)$RM which, for a frequency $\nu=1$ GHz, gives $d\chi/d\nu\approx -0.01\times$RM deg-MHz$^{-1}$ when RM is expressed in units rad-m$^{-2}$. There are several contributions to RM: the stellar magnetosphere, the local interstellar medium, the interplanetary medium, and the ionosphere. The latter three contributions sum to a value of order unity. The stellar value RM$_\star=2.89\times 10^{-3}\int n_e(s)B_{||}(s) ds$ rad-m$^{-2}$, where $s$ is in units of $R_\star$. Again assuming a mass loss rate per unit area similar to that of the Sun and setting $B_{||}\sim|B|\cos\theta$ we have RM$\sim 60$ for reflections occurring at 5-7 $R_\star$, yielding a change in the EVPA across the 200~MHz bandwidth of 120$^\circ$. Even with poor SNR a change this large would be easily seen. The change is no more than 10$^\circ$ ($1\sigma$), suggesting that RM$_\star$ has been over-estimated by an order of magnitude or more, implying that the density is significantly lower than assumed. This may be the result of a mass loss rate per unit area that is $\sim\!10\%$ that of the Sun. 

As discussed in \S4.3 the over-dense plasma structure that exists some distance from the star may be an equatorial plasma disk maintained by centrifugal force between $r_i$ and $r_o$, and the range of densities that could occur in such a disk lie within the range needed for reflection. However, as demonstrated by \citet{UdDoula2008} there may be no steady-state disk possible -- that it is, in fact, dynamic. In the case of UV~Cet, disk dynamics might determine the occurrence, the phase, the repeatability, and the life times of linearly polarized reflections. Hence, they are potentially valuable diagnostics of over-dense structures that could be studied with dynamic spectropolarimetry performed over a range of time scales. 


\section{Concluding Remarks}

MeerKAT enables sensitive dynamic spectroscopy of stellar radio emission over significant frequency bandwidths. The radio outburst from UV~Cet observed in October 2021 produced detailed dynamic spectra in all four Stokes parameters. These are consistent with the emission of x-mode radiation near the electron gyrofrequency amplified by the CMI mechanism. We have presented a schematic model in which the CMI radiation originates on active field lines in a stellar magnetosphere. The magnetosphere is modeled as an axisymmetric dipole that is aligned with the rotational axis of the star, which is taken to be at an inclination of $60^\circ$. The CMI produces radiation that is beamed into the walls of a hollow cone. Hence, CMI radiation is only beamed into the line of sight when a source on a particular field line at the appropriate resonant height is at particular longitudes. We find that the arc-like substructures in peaks A, B, and C can be understood in terms of bursty emission from AFLs spanning a range of longitudes rotating toward or away from the observer. 

We have made no attempt to produce a fully self-consistent model of the CMI emission from UV~Cet. To do so requires better understanding of the magnetic field and the plasma environment around the star. We considered UV~Cet as a rigidly rotating magnetosphere, which may lead to the presence of an over-dense equatorial disk for $r>r_i\approx 4$ R$_\star$. Instabilities in the equatorial disk can result in mass infall and mass breakout \citep{UdDoula2008}; interaction of disk material with the magnetosphere may provide the quasi-persistent source of fast electrons in active longitudinal bands needed to drive the CMI. While peaks A and C may be consistent with CMI emission from magnetic field lines that map into the inner edge of an equatorial disk ($L\sim 4$), we cannot account for the apparent frequency drift rates of arcs in peak B and some of the arcs in peak C unless the emission originates from AFLs with $L\sim 2$. Since the resonant heights for the model dipole are only $\sim 1.3-1.5$ R$_\star$ and since $\langle B\rangle=4$ kG, a pure dipole representation of the magnetic field may break down at these heights as higher-order multipole terms contribute.  \citet{Barnes2017} find that the distribution of photospheric spots peaks between latitudes of $40-70$ deg with no evidence for a dominant polar spot. Hence, while the simple model presented here can account for spectral features in terms of the large scale magnetic terms, it only does so qualitatively. A more realistic model of the stellar magnetic field is needed. 

We have considered the plasma conditions under which amplification of RCP emission near the electron gyrofrequency can occur via the CMI mechanism and the conditions necessary for its escape. We find that it can be amplified and escape the source without being absorbed at the second harmonic layer for an ambient plasma density of $1-5\times 10^6$ cm$^{-3}$ in the source. It seems likely that the RCP emission is partially depolarized as it propagates to the observer, possibly as the result of mode coupling in a QT region. A remarkable aspect of the observed CMI emission is the presence of elliptically polarized components during the declining phase of peak C. While the elliptical polarization may be intrinsic to the source rather drastic limits must be imposed on the plasma density near the source. We have suggested that the linearly polarized burst components may be the result of the reflection of RCP emission from an over-dense plasma structure.

It is interesting to note that \citet{Zic2019}, who observed UV~Ceti for two successive rotational periods on 2018 Oct 2 and again on 2019 Mar 7, found that substructures in the outbursts during a given epoch repeated in successive rotations, and found similar drift rates of substructures even between epochs, which were separated by nearly 700 stellar rotation periods. The substructures observed in the MeerKAT spectra differ in significant ways from those observed by ASKAP (although comparisons must account for differences in sensitivity and bandwidth between the two telescopes). Clearly, the physical processes driving the radio outburst each rotation period persist for at least $\sim 40$ ksec (2 rotation periods) and may persist for 100s of rotation periods but ultimately evolve in time to produce changing emission profiles. On the other hand, the polarity of the large scale magnetic field, as evidenced by the dominance of RCP radio outbursts at decimeter wavelengths, appears to persist for decades. 

The radio outbursts from UV~Cet raise a number of fascinating puzzles. While the outbursts are consistent with CMI radiation from a stellar magnetosphere at the fundamental of the electron gyrofrequency in the x-mode, many questions remain unresolved. These include:

\begin{itemize}
\item What is the detailed structure of the stellar magnetic field? How does it evolve in time? Why does it differ so dramatically from that of BL~Cet?
\item What is the distribution and temperature of plasma around the star? How does it evolve in time?
\item What is the mass loss rate from the star in the form of a stellar wind? How is it distributed in three dimensions?
\item By what process(es) are the fast electrons produced that drive the CMI? And the energetic electrons that produce nonthermal gyrosynchrotron emission, both in quiescence and in flares?
\end{itemize}

Looking forward, observations are needed on a broad front. Higher resolution ZDI observations of the magnetic field are needed to constrain the structure of the magnetosphere. Broader bandwidth spectra are needed to observe the full range of heights over which the CMI is operative. A hope is that such observations can be performed contemporaneously as the basis for a joint reconstruction of the stellar magnetosphere. Observations at higher radio frequencies, where CMI emission is no longer present, are needed to characterize the nonthermal gyrosynchrotron emission and flare activity on the star. Observations in the O, UV, and X-ray bands are needed to constrain the temperature and distribution of plasma around the star and to further characterize activity related to auroral emission and flares. The fact that the magnetosphere and magnetospheric processes like the CMI evolve in time imposes significant challenges on coordinating joint observations, however. Finally, such observations need to be expanded to comprise a significant sample of fully convective late-type dwarfs to identify those that are auroral emitters. These stars represent a special class of objects that share attributes of both planets like Jupiter, and stars like the Sun. 

\begin{acknowledgments}
The MeerKAT telescope is operated by the South African Radio Astronomy Observatory, which is a facility of the National Research Foundation, an agency of the Department of Science and Innovation.The National Radio Astronomy Observatory is a facility of the National Science Foundation operated under cooperative agreement by Associated Universities, Inc. GH acknowledges support provided by the Simons Foundation grant "Planetary Context of Habitability and Exobiology." We thank the referee for their careful reading of the manuscript and constructive comments. 
\end{acknowledgments}

\bibliography{UVCet.bib}{}
\bibliographystyle{aasjournal}

\end{document}